\documentclass[11pt]{article}
\usepackage{amsmath,amssymb,amsfonts,epsfig}

\newtheorem{prop}{Proposition}[section]

\newtheorem{rem}[prop]{Remark}

\newcommand{\proof}[1]{\vspace{1.5ex}{\small 
{\bf Proof:} #1 \hfill$\blacksquare$}\vspace{1.5ex}}

\newcommand{\beq}{\begin{equation}}
\newcommand{\eeq}{\end{equation}}

\newcommand{\beqa}{\begin{eqnarray*}}
\newcommand{\eeqa}{\end{eqnarray*}}

\newcommand{\beqan}{\begin{eqnarray}}
\newcommand{\eeqan}{\end{eqnarray}}

\newcommand{\supp}{\protect{{\rm supp\;}}}

\newcommand{\omsh}[1]{\protect{\omega_{\bf #1}}}
\newcommand{\omshi}[3]{\protect{\omega_{{\bf
#1}_{#2}^{#3}}}} 

\newcommand{\ra}[1]{\protect{{\bf #1}}}

\newcommand{\mc}[1]{\mathcal{#1}}

\newcommand{\iu}{{\rm i}}

\newcommand{\R}{\mathbb R}

\title{The ultraviolet infrared mixing problem on the noncommutative Moyal space}

\author{Dorothea Bahns
\\
\\

\begin{tabular}{l}
{\footnotesize Courant Research Centre ``Higher Order Structures in Mathematics''}\\
{\footnotesize Universit\"at G\"ottingen,  Bunsenstr. 3-5,  D - 37073 G\"ottingen, Germany}
\end{tabular}
\\
{\footnotesize 
 {\tt bahns@uni-math.gwdg.de}}
}

\date{}

\begin{document}

\maketitle

\begin{abstract}
\noindent 
It is shown that the mixing of ultraviolet and infrared divergences in quantum field theory on Moyal space is not an artefact of the Euclidean framework, but occurs also in the Hamiltonian setting when the interaction is given in terms of the Moyal twisted convolution product. The mixing mechanism in both settings is examined from the point of view of microlocal analysis and it is shown that they are different from one another.

\end{abstract}




A number of thought experiments and arguments from theoretical physics indicate that the geometry of spacetime at very small scales might not be smooth. In particular, it is believed  
that there should be restrictions on how well an event in spacetime can be localized. In~\cite{DFR} it was proposed that such restrictions take the form of uncertainty relations, such that one might localize very well in some directions, at the cost of losing precision in the others. These relations were then realized  -- in the spirit of quantum mechanics -- by replacing coordinates by non-commuting operators, which in the simplest case ('Moyal space') satisfy canonical commutation relations. Physical consequences of such a modification of the spacetime structure should be visible in particle physics experiments, so it is crucial to understand quantum fields on such noncommutative spaces.

Most of the literature on quantum fields on the noncommutative Moyal space concerns  the framework of the modified Feynman rules~\cite{filk}, partly because these are also motivated by string theory~\cite{schomerus}. In this framework, one takes Euclidean quantum field theory, which is based on an elliptic partial differential operator, as a starting point, and replaces every local product of fields by a twisted convolution product, the so-called Moyal product. The most prominent feature of this framework is the so-called ultraviolet-infrared mixing problem found in~\cite{uvir}:  some contributions to the perturbative expansion which are by themselves regular by virtue of the Moyal product, turn out to be ill-defined when a number of them appears within more complicated contributions. This effect renders scalar $\varphi^n$ field theories non-renormalizable, unless one adds the so-called Grosse-Wulkenhaar term which modifies the propagator~\cite{GrW,R}. 

In ordinary quantum field theory on vector spaces, the Euclidean framework is a helpful tool. Certain calculations are simpler in this framework than in the physically meaningful Minkowskian setting, which is based on a hyperbolic partial differential operator, and a  correspondence between the two settings guarantees that the Euclidean  calculations can be transferred to the Minkowskian realm. On Moyal space, on the other hand,  there is no simple link between Euclidean and Minkowskian settings. A naive extension of the modified Feynman rules to a Minkowskian setting leads to a violation of unitarity~\cite{gomis} -- and vice versa, starting from a unitary Minkowskian theory, it is not clear whether a consistent  Euclidean counterpart can be found~\cite{bahns_schwinger}.
Some progress was made recently in understanding the Grosse-Wulkenhaar term in a hyperbolic setting~\cite{fischSzab}, but a number of technical problems still remain to be solved and the resulting theories seem to be plagued by strange divergences~\cite{zahn}. 

With the connection between the Euclidean and Minkowskian  realm obscure, and since first calculations indicated that massive theories of hyperbolic signature might be renormalizable, it was thought for a while that -- while the infrared regime is indeed drastically modified~\cite{qplan1} -- the ultraviolet-infrared mixing as such might be absent in such theories~\cite{fischer}. We will, however, see that in the Hamiltonian framework with an interaction term given by the Moyal product\footnote{There is a certain amount of freedom we have in the definition of interaction terms on Moyal space. It is not clear yet, whether the averaged Hamiltonian first proposed in~\cite{DFR} or the ultraviolet finite interaction term proposed in~\cite{BDFP_quantumdiag} lead to a mixing problem as well.}, a mixing does occur -- albeit by a different mechanism than the one found in the setting of the modified Feynman rules.

The paper is organized as follows:  After explaining some notation and the general setup, some tools from microlocal analysis are recalled, most notably the notion of the wavefront set and the singular order of distributions as well as their relation to renormalization theory. In the third section, the mixing mechanism in Euclidean theories on Moyal space is explained in terms of (Fourier transforms of) distributions and their wavefront sets, and it is shown that the corresponding terms in the Minkowskian theory (Hamiltonian setting) do not necessarily show this mixing. In the subsequent section, the paper's main point is made: Using techniques of microlocal analysis, it is shown that a mixing of a different kind does occur in certain graphs in the hyperbolic framework -- a problem which in turn is not present in the Euclidean framework. This last section can be read independently of section \ref{sec:EuMix} which requires some familiarity with Feynman graphs.


\section{Setup and Motivation}

We consider a scalar massive field with polynomial self-interaction on a noncommutative space whose coordinates are subject to commutation relations of the form 
\[
\, [ x^\mu,x^\nu] = i \theta^{\mu \nu} \qquad \mu,\nu=0,\dots,d-1
\]
with an antisymmetrix real $d\times d$-matrix $(\theta^{\mu\nu})$ of maximal rank (Moyal space). In the simplest setup, such models can be understood as models on ordinary $\R^d$ (Euclidean setting) or its Minkowskian counterpart $\R^{1,d-1}$ (hyperbolic setting), but with a nonlocal interaction term given by a twisted convolution product. 
Various approaches to derive a perturbative expansion for such an interaction term have been worked out. Here, we will consider the setting of the modified Feynman rules (Euclidean setting) and the Hamiltonian formalism (Minkowskian regime).

In ordinary field theory, the analytic expressions (nested integrals, distributions etc.) produced by the perturbative setup can be encoded in terms of Feynman graphs, which are made up of (arbitrarily many) vertices to which a certain number of edges can be attached ($n$ in $\varphi^n$-theory). An edge can be either open, i.e. be attached to one vertex with only one of its ends (``external leg''), or it can connect two different vertices, or it can start and end at the same vertex. An edge connecting two different vertices corresponds, in the analytic expression, to a fundamental solution of the partial differential operator $P$ governing the theory. For the models under consideration, we have $P=-\Delta + m^2$ in the Euclidean setting, with $\Delta$ denoting the Laplace operator, and $m>0$ the theory's mass parameter,  and $P=\partial_{x_0}^2 -\Delta_{\ra x}+ m^2$ in the hyperbolic setting, with $(x_0,\ra x) \in \R\times \R^{d-1}$ and $\Delta_{\ra x}$ denoting the Laplace operator on $\R^{d-1}$. In the former case, the fundamental solution is of course unique, and in the latter case, the fundamental solution which appears in the perturbative expansion turns out to be the Feynman propagator.

When a twisted convolution product appears in the interaction term, perturbation theory is more complicated~\cite{filk, bahns_diss, sibold}. Generally, the (Fourier transforms of the) analytic expressions produced by perturbation theory contain so-called twisting factors that depend on momenta $p_1, \dots,p_n \in \R^d$,
\[
e^{-\frac \iu 2 \sum_{i<j} \langle p_i,  \theta p_j \rangle }
\]
where $\langle \cdot, \cdot \rangle$ denotes the scalar product on $\R^d$ or the inner product on $\R^{1,d-1}$, respectively, and $\theta$ is the maximal rank antisymmetric matrix from the commutation relations. We assume $d$ to be even, such that $\det \theta \neq 0$.

Such twisting factors may cancel or add up in a given contribution to the perturbative expansion. 
In a graphical language this can be encoded by keeping track of the order in which edges connect different vertices: Compared to an ordinary Feynman graph, we replace each vertex by a number of dots ($n$ in $\varphi^n$-theory) and keep track of which of these dots are connected by edges. In this manner, we find many contributions which make up what would ordinarily be only one graph. For instance, two (of many) contributions to the ordinary fish graph 
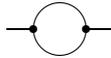
\[
\begin{picture}(60,10)(-10,0)
 \put(5,0){\circle*{3}}
  \put(25,0){\circle*{3}}
  \put(15,0){\circle{20}}
  \put(-5,0){\line(3,0){10}} 
  \put(25,0){\line(3,0){10}} 
\end{picture}
\]
are  
\[
\begin{picture}(60,10)(-5,0)
\put(0,2){\circle{2}}
\put(10,2){\circle{2}}
\put(20,2){\circle{2}}
\put(30,-2){\line(0,1){5}} 
\put(40,2){\circle{2}}
\put(50,2){\circle{2}}
\put(60,2){\circle{2}}
\qbezier(10,2)(30,18)(50,2)
\qbezier(20,2)(40,18)(60,2)
\end{picture} 
\qquad \mbox{ and }  \qquad  
\begin{picture}(60,10)(-5,0)
\put(0,2){\circle{2}}
\put(10,2){\circle{2}}
\put(20,2){\circle{2}}
\put(30,-2){\line(0,1){5}} 
\put(40,2){\circle{2}}
\put(50,2){\circle{2}}
\put(60,2){\circle{2}}
\qbezier(10,2)(35,23)(60,2)
\qbezier(20,2)(35,15)(50,2)
\end{picture} 
\]
Here, the small vertical line separates the two vertices.
Attaching an open edge to those dots in a vertex with no edge attached
and shrinking all the dots of each vertex to one point,  we would recover  the ordinary fish graph in both cases, but on Moyal space,  the two contributions of the perturbative expansion which correspond to these graphs differ from one another. Appendix~\ref{app:rules} contains the rules how to recover the analytic expressions of the perturbative expansion from such graphs in the Euclidean and the Hamiltonian Minkowskian setting. In particular, one finds that in the Euclidean realm, edges still correspond to the fundamental solution of the model's elliptic partial differential operator, while in the presence of twistings, in the hyperbolic theory, an edge in general no longer corresponds to the Feynman propagator\footnote{The reader who is familiar with the modified Feynman rules might think this graphical language impractical, since it makes it difficult to identify graphs which lead to the same  expression, or to identify, say, 1-particle-irreducible graphs. However, note that we will partly work with all expressions still comprising testfunctions (i.e. before performing the adiabatic limit, see below), in which case the modified Feynman rules do not lead to expressions that are  invariant under cyclic permutations of the dots of each vertex. Moreover, in the hyperbolic setting, this graphical language is the most efficient one which is known, and so we shall employ it in both settings, to make the comparison between the two approaches easier.}.

We will in general proceed as follows: given a graph, we use the rules listed in appendix~\ref{app:rules} to write down the corresponding, at this stage only formal analytic expression -- typically, such an expression is given as an integration over a (non-integrable) function and some oscillating factors. We then make sense of such an expression in terms of distributions. The following notation is used throughout:

\medskip 

{\bf Notation:} We write $kx$ and $k\theta p$ for inner products $\langle k, x \rangle$ and $\langle k, \theta p\rangle$. In particular, when we work in the Minkowskian setting, we write $x=(x_0,\ra x)$ and $p=(p_0, \ra p) \in \R\times \R^{d-1}$, and use an expression like $p^2$ as short-hand notation for  $p_0^2-\ra p^2$ with $\ra p^2$ denoting the scalar product of $\ra p$  with itself. A tilde on a vector $p\in \R^d$ means that it is on the positive mass shell, $\tilde p= (\omega_{\ra p}, \ra p)$ where $\omega_{\ra p}=\sqrt{\ra p^2 + m^2}$. A tilde on a function or a distribution denotes its Fourier transform.


\medskip 

As a simple example let us now consider one of the  contributions to the ordinary tadpole graph 
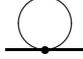
\[
\begin{picture}(30,-10)
\put(15,2){\circle*{3}}
\put(0,2){\line(1,0){30}} 
\put(15,12){\circle{20}}
\end{picture} 
\]
in $\varphi^4$-theory on Moyal space, the nonlocal tadpole
\beq\label{fig:tadpole}
\begin{picture}(30,10)
\put(0,2){\circle{2}}
\put(10,2){\circle{2}}
\put(20,2){\circle{2}}
\put(30,2){\circle{2}}
\qbezier(10,2)(20,21)(30,2)
\end{picture} 
\eeq
According to the rules from the appendix, the, for now formal, expression which corresponds to this graph is
\[
\int \frac 1 {p^2+m^2} \ e^{- \iu (q^\prime+q)x} \ 
e^{-\frac \iu 2 q^\prime \theta q- \iu  p \theta q }  \ g(x) 
\ d p\, dx 
\]
where $g$ is a testfunction, $g \in \mc D(\R^d)$. This expression can be understood in terms of formal integral kernels of distributions as
\beq\label{eq:tadpoleEuNoAL}
\tilde g(q^\prime+q) e^{-\frac \iu 2 q^\prime \theta q} \ G_E(\theta q)
\eeq
where
\beq\label{eq:GE_OsciInt}
G_E(x)= \int \frac 1 {p^2+m^2} \  e^{- \iu  p x } \; d p
\eeq
is the fundamental solution of $-\Delta +m^2$, written as an oscillatory integral. We will see in the next section that $G_E(\theta q)$ can be understood as the pullback (in the sense of distributions) of $G_E$ along the linear map $\theta$ and indeed is a distribution on $\R^d$ by the non-degeneracy of~$\theta$.

The testfunction $g$ which appears here has an interpretation in terms of physics:  it restricts the interaction region to a compact region in spacetime. 
At least in the usual framework, however, physical quantities are calculated in what is called the adiabatic limit -- where these testfunctions approach a constant (the coupling constant). In~\cite{epstein}, this limit was investigated in  terms of the appropriate topology, but for theories on Moyal space, no consistent picture exists as yet~\cite{zahn_diss}. In order to be able to make our point, we shall disregard these problems for now, and use a naive version of the adiabatic limit, where at some point, we set all testfunctions equal to 1, or equivalently, replace their Fourier transforms by $\delta$-distributions. Observe that this is done throughout the literature on quantum field theory in the Euclidean framework without mention.

In this naive adiabatic limit, we find for the nonlocal tadpole the well-known expression
\beq\label{eq:tadpoleEu}
\delta(q^\prime+q) e^{-\frac \iu 2 q^\prime \theta q} \ G_E(\theta q)
\eeq
Observe that the $\delta$-distribution means that effectively, the  twisting factor involving only the external momenta $q, q^\prime$ will give 1, by the antisymmetry of $\theta$, and could be dropped from the expression. 

It seems hopeless to understand the formal expressions produced by the rules without such simplifications, and  since this is what is done in the Euclidean framework without further mention, we will, generally, use the $\delta$-distributions which are produced by the adiabatic limit to simplify twistings as follows:

{\rem \label{rem:adiabTrick}\rm Let $u$ be a distribution on $\R^n$ such that
\[
u(g)=\int  h(k_1,\dots ,k_m) \, \tilde g(k_1 + \dots + k_m) \, dk_1 \cdots dk_m
\]
with a function $ h$ on $\R^{nm}$. Then, the adiabatic limit, where $g$ is replaced by a constant, produces  a $\delta$-distribution  $\delta(\sum k_j)$. If $h$ contains twisting factors, we use this $\delta$-distribution to simplify them as far as possible. Denoting the resulting function by $h_{simp}$, we then consider the distribution 
\[
u_{simp}(g)=\int h_{simp}(k_1,\dots ,k_m) \, \tilde g(k_1+\dots + k_m) \, dk_1 \cdots dk_m
\]
with the testfunction $g$ intact again. 
}



\section{Some microlocal analysis}

Let us recall some general mircolocal techniques. First observe that 
a compactly supported distribution whose Fourier transform quickly decreases in all directions is smooth. The wavefront set is designed to take this property into account, and for a general distribution $u\in \mc D^\prime (\Omega)$ with $\Omega$ open in $\R^n$, it encodes not only the singular support, but also information on the directions in which its Fourier transform does not  quickly decrease -- in the language of physics: it not only encodes the distribution's position space singularities, but also its Fourier transform's  behaviour at infinity (at large momenta, i.e. in the 'ultraviolet' regime), which causes the singularities. The wavefront set  
$WF(u)$ of a distribution $u$ on $\Omega \subseteq \R^n$ is therefore a subset of the cotangent space with $0$ removed, $\Omega\times \dot \R^n$, where $\dot \R^n = \R^n \setminus \{0\}$, and the projection to the first factor (base point) is the singular support of $u$. See~\cite{hoerm} for details. For example, the wavefront set of the $\delta$-distribution on $\R^n$ is 
\[
WF(\delta)=\{(0;p) \in \R^n \times \dot \R^n \ | \ p \neq 0\}
\]
Frequently, redundant notation as above will be used, where the fact that $p$ is non-zero is emphasized (although it is clear, for $p$ is in $\dot \R^n$). 
Since $G_E$ is the fundamental solution of an {\em elliptic} partial differential operator, by  elliptic regularity, its wavefront set must be contained in that of the $\delta$-distribution, and it can be shown, for instance, using the representation (\ref{eq:GE_OsciInt}) of $G_E$ as an oscillatory integral that $WF(G_E)$ is in fact equal to $WF (\delta)$,
\[
WF(G_E)  = \{ (0;p) \in \R^n\times \dot \R^n \} 
\]
In particular, the singular support of $G_E$ is $\{0\}$. It will be crucial in renormalization theory to be able to quantify how 'bad' this singularity is. The suitable notion to do so is Steinmann's scaling degree (in 0), defined for any distribution $u \in \mc D^\prime(\R^n)$ as
\[
scal(u) = \inf\{s\in \R \ |  \ \lim_{\lambda \searrow 0} \lambda^s u(g_\lambda)=0 \mbox{ for all testfunctions } g \} 
\]
where $g_\lambda(x)=\lambda^{-d}g(\lambda^{-1}x)$ for $\lambda > 0$.
The scaling degree of $\delta\in \mc D^\prime(\R^d)$ is $d$, and  the representation of $G_E$ as an oscillatory integral (\ref{eq:GE_OsciInt}) reveals that the scaling degree of  $G_E$ on $\R^d$ is $d-2$. 

The notion of the scaling degree can be extended to distributions which are defined only on the open set $\dot \R^n \subset \R^n$.
If the scaling degree of  $u \in \mc D^\prime(\dot \R^d)$ is  less than the dimension $d$, $u$ uniquely extends to a distribution in $\mc D^\prime(\R^d)$ of the same scaling degree, while 
a scaling degree equal to or larger than the dimension $d$ means that $u$ can be extended but the extension is not unique.
With the {\em singular order} of a distribution defined as the difference of scaling degree and dimension, $scal(u) - d$, this means that a distribution extends uniquely if it has negative singular order, and non-uniquely otherwise.
 In the non-unique case,  the extension of the distribution is given, in the language of physics, in terms of subtractions of {\em counterterms}, and the non-uniqueness corresponds to the fact that there is a certain amount of freedom in the choice of the counterterms (finite renormalizations).  Generally, the counterterms are given in terms of derivatives of the $\delta$-distribution, and as such are considered to be {\em local}. See \cite{BrFred} for details, and appendix~\ref{app:renorm} for the extension map and an example.

We now recall a special case of Thm. 8.2.4 in \cite{hoerm}. Let $A: \R^n\rightarrow \R^m$ be a linear map, then a distribution $u \in \mc D^\prime(\R^m)$ can be pulled back along $A$, if the set of normals of $A$,
\[
N_A=\{ (A x,k) \in \R^m\times \R^m\ | \ A^t \, k = 0\}
\]
has empty intersection with $WF(u)$. This pullback $A^*u \in \mc D^\prime(\R^n)$ uniquely extends the pullback of smooth distributions, $A^*u=u \circ A$, so we formally write $u\circ A$ also when using formal integral kernels. The pullback's  wavefront set is contained in 
\beq\label{eq:WF_Au}
WF(A^*u) \subseteq
A^* (WF(u)) = \{(x,A^t \, k) \in \R^n\times \R^n \ | \ (A x,k) \in WF(u)\} 
\eeq
If $A$ is non-degenerate, we have $N=\{ (x,0) \in \R^m\times \R^m \}$, so any distribution on $\R^m$ can be pulled back along $A$. 

This theorem can be used in particular to explain the product of distributions as the pullback of the tensor product $u \otimes v$ along the diagonal map ${\rm diag}(x)=(x,x)$. The set of normals of ${\rm diag}$ is $\{(x,x;k,p)|k+p=0\}$, so the product of two distributions $u,v \in \mc D^\prime (\R^n)$ is a distribution in $\mc D^\prime (\Omega)$, $\Omega \subseteq \R^n$ open, if $(x,p)\in WF(u)$ implies $(x,-p) \notin WF(v)$  for all $x \in \Omega$. We will call this condition on the wavefront sets H\"ormander's criterion. Observe that this is not an ``only if'' condition. The wavefront set of the resulting distribution 
is contained in the set
\beqan\label{eq:WFprod}
&&\big( \ \{(x,p)\, | \, (x,p)\in WF(u) , \,  x\in \supp v \}  \ \cup  \ 
\{(x,p)\, | \, (x,p)\in WF(v) ,  \, x\in \supp u \}  \nonumber \\
&&\quad \cup  \ 
\{(x,k+p)\, | \, (x,k)\in WF(u) ,   \, (x,p)\in WF(v) \} \ \big) \cap \Omega \times \R^n 
\eeqan
The scaling degree of a product of distributions is less than or equal to the sum of the individual factors' scaling degrees.

As an example, consider the fundamental solution $G_E$. Monomials $G_E^k$ are distributions in $\mc D^\prime(\dot \R^d)$, since the singular support of $G_E$ is $\{0\}$. H\"ormander's  criterion is, however, clearly not satisfied in 0. The scaling degree of $G_E^k \in \mc D^\prime(\dot \R^d)$ in 0 is $k\cdot (d-2)$, so, for $d \geq 4$, a $k$-fold product ($k \geq 2$) cannot be uniquely extended to a distribution in $\mc D^\prime(\R^d)$, but always requires renormalization. For $d=3$, only the 2-fold product can be extended uniquely, and for $d=2$, none of the products need to be renormalized.
To establish  contact with the formulation of quantum field theory in momentum space, observe that the singular order of a distribution
correpsonds to the power counting degree of divergence in momentum space~\cite{BrFred}, e.g. the singular order of $G_E^2$ is $d-4$, and therefore agrees with the power counting degree of divergence of its formal Fourier transform $\widetilde{G_E} \times \widetilde{G_E}$,
\[
 \int \frac 1 {(p-k)^2+m^2} \frac 1 {k^2+m^2} \ dk
\]

The Feynman propagator $G_F$, on the other hand, has singular support on the boundary of the lightcone, but its wavefront set is such that outside $x=0$, all the cotangent vectors on the positive cone and all those on the negative cone point in the same direction. Therefore,  H\"ormander's criterion is satisfied for $x\neq 0$ and we find that $G_F^k \in \mc D^\prime(\dot \R^d)$ as in the elliptic case. Observe that this is not true for the advanced and retarded solution. In 0, H\"ormander's criterion is not satisfied, and as the scaling degree for the Feynman propagator is the same as that of $G_E$, we find the same need for renormalization as in the elliptic case. 
Last not least, it should be noted that the perturbative expansion not only produces products of fundamental solutions, but also convolution products; in terms of formal integral kernels, a typical contribution would be  $G_F(x_1-x_2)^2 G_F(x_2-x_3)^2 G_F(x_1-x_3)$ (analogously with $G_E$ in the Euclidean setting). By H\"ormander's criterion, such distributions are elements of $\mc D (\R^{kd}\setminus D)$ with $k=3$ in the example, where $D$ denotes the `fat' diagonal of pairwise (or more) coincidences,  $D=\{(x_1,\dots,x_k) \in \R^{kd}\ | \ x_i=x_j \mbox{ for some } i\neq j\}$. Renormalization theory is concerned with extending such distributions to $\R^{kd}$. 
As the form of the counterterms changes with growing scaling degree, one needs an infinite number of types of counterterms, if arbitrarily high scaling degrees occur. In this case we say that the theory is not (locally) renormalizable.

Let us now reconsider the tadpole (\ref{eq:tadpoleEu}) on Moyal space in this language. The formal expression $G_E \circ \theta$ from (\ref{eq:tadpoleEu}) can be understood as the pullback $\theta^* G_E\in \mc D^\prime (\R^d)$ along the non-degenerate linear map $\theta:\R^d \rightarrow \R^d$, and from (\ref{eq:WF_Au}) we deduce that
\beq\label{eq:WF_GEtheta}
(WF(\theta^* G_E)) =\{(x,\theta^t \, k) \in \R^d\times \R^d \ | \ (\theta x,k) \in WF(G_E)\}  = WF(G_E)
\eeq
Also, $\theta^*G_E$ has the same scaling degree as $G_E$. The peculiarity about the tadpole  (\ref{eq:tadpoleEu}) is that $\theta^*G_E$ appears as a distribution on {\em momentum} space. Therefore, the ultraviolet divergence of $G_E$ in 0 now occurs  at small momenta, i.e. in the infrared region. As we shall see below, this will lead to curious divergences when tapoles (\ref{eq:tadpoleEu}) are inserted as subgraphs in a larger graph (ultraviolet-infrared-mixing problem).



\section{The mixing on Euclidean Moyal space}\label{sec:EuMix}

We will now state the ultraviolet-infrared mixing problem in the language of distributions and  wavefront sets. To understand the underlying mechanism, let us first consider a graph from {\em ordinary} quantum field theory which contains a line-like subgraph, i.e. a graph of the form 
\beq\label{fig:genertildeu}
\cdots \hspace{4ex} \underbrace{\begin{picture}(120,10)(10,18)
  \put(0,20){\circle*{3}}
  \put(20,20){\circle*{3}}
  \put(40,20){\circle*{3}}
  \put(60,20){\circle*{3}}
  \put(72,16){$\cdots$}
  \put(100,20){\circle*{3}}
  \put(130,20){\circle*{3}}
  \put(150,20){\circle*{3}}
  \put(-5,10){$x_0$}
  \put(15,10){$x_1$}
  \put(35,10){$x_2$}
  \put(55,10){$x_3$}
  \put(96,10){$x_{r\!-\!1}$}
  \put(125,10){$x_r$}
  \put(145,10){$x_{r+1}$}
  \qbezier[12](0,20)(10,30)(20,20)
  \qbezier[12](0,20)(10,10)(20,20)
  \qbezier[12](20,20)(30,30)(40,20)
  \qbezier[12](20,20)(30,10)(40,20)
%
  \qbezier[12](40,20)(50,30)(60,20)
  \qbezier[12](40,20)(50,10)(60,20)
%
%
  \qbezier[7](60,20)(63,25)(68,25)
  \qbezier[7](60,20)(63,15)(68,15)
  \qbezier[7](100,20)(97,25)(92,25)
  \qbezier[7](100,20)(97,15)(92,15)
  \qbezier[12](100,20)(115,30)(130,20)
  \qbezier[12](100,20)(115,10)(130,20)
  \qbezier[12](130,20)(140,30)(150,20)
  \qbezier[12](130,20)(140,10)(150,20)
  \put(27,30){$u_1$}
  \put(47,30){$u_2$}
  \put(107,30){$u_{r-1}$}
  \put(7,30){$u_0$}
  \put(137,30){$u_r$}
  \end{picture} \phantom{\int_z}}_{\mbox{no open edges}}
  \quad \ \cdots 
\eeq
Here,  $x_0,\dots,x_{r+1}$ label the vertices, and for $j \in \{0, \dots, r\}$, $u_j$  labels the formal integral kernel of the distribution which corresponds to the little subgraph between vertex $x_j$ and $x_j+1$ (dotted lines). 
Observe that these little subgraphs may in general contain more vertices, e.g. in $\varphi^3$-theory, we might have
\beq\label{fig:complicSubgr}
\begin{picture}(40,35)(10,3)
  \put(0,30){\circle*{3}}
  \put(0,10){\circle*{3}}
  \put(-5,20){\circle*{3}}
  \put(20,20){\circle*{3}}
%
  \put(-16,12){\small{$x_{j}$}}
  \put(21,12){\small{$x_{j+1}$}}
\put(-20,17){\small{$\cdots$}}
\put(23,17){\small{$\cdots$}}
  \qbezier(0,30)(15,35)(20,20)
  \qbezier(0,10)(15,5)(20,20)
  \qbezier(0,30)(-10,20)(0,10)
  \qbezier(0,10)(10,20)(0,30)
\end{picture} 
\eeq
whose corresponding analytic expression is of the form 
\[
u_j(x_j-x_{j+1}) = \int v(z,z^\prime,x_j-x_{j+1})\ g(z) \, g(z^\prime) \ dz\, dz^\prime
\]
with evaluations in additional testfunctions associated to the additional vertices $z$ and $z^\prime$. Here,  $v$ is a distribution given in terms of convolutions of the fundamental solution $G_E$ with itself, which  in fact requires renormalization.

We will, however, for now assume that the analytic expressions $u_j$ which correspond to the little subgraphs, as well as their products as they occur in (\ref{eq:conv}) below,  are well-defined distributions. This will be justified later by the examples we study.
The analytic expression corresponding to the full graph (\ref{fig:genertildeu}) then is of the form
\beq\label{eq:conv}
\int 
w(x_0,x_{r+1}) \ \prod_{i=0}^{r} u_i(x_i-x_{i+1})  \ 
g(x_0) \cdots g(x_{r+1})
\
d x_0 \cdots d x_{r+1}
\eeq
where $w$ is a distribution which encodes the analytic expression for the part of the graph which remains entirely unspecified -- it contains evaluations in additional testfunctions $g$ corresponding to further  vertices.  The distributions $u_i$ depend only on relative coordinates. Observe that this is a consequence of the condition that there are no open edges attached to the vertices $x_1, \dots, x_r$.

 We rewrite  (\ref{eq:conv}) in terms of the Fourier transforms of $w$ and the $u_j$,
\beq\label{eq:tildeginlinegr} 
\int 
\tilde w(q,p)  \ \prod_{i=0}^{r} \tilde u_i(p_i)  \ 
\tilde g(q+p_0)  \tilde g(p-p_r)  \prod_{j=1}^{r} \tilde g(-p_{j-1}+p_j)  
\ d p_0 \cdots d p_r 
\; dq\, dp
\eeq
Observe here, that in ordinary massive Euclidean quantum field theory, those Fourier transforms are generally smooth.
Now, in the adiabatic limit, $\tilde g$ is replaced by the $\delta$-distribution, so we then find
\beqan\label{eq:line_moment}
\int 
\tilde w(-p,p) \; \tilde u_0(p)\; \tilde u_1(p)\;  \cdots \; \tilde u_{r-1}(p)\; \tilde u_r(p) \  dp
\eeqan
Observe that our assumption on the existence of the products $\prod u_j(x_j-x_{j+1})$  translates here to the assumption that the integrand in (\ref{eq:line_moment}) decreases quickly enough at large momenta for the integral to exist.


\subsection{Insertions of nonlocal tadpoles}

It was shown in~\cite{uvir} that the nonlocal tadpole graph (\ref{fig:tadpole}) produces an infrared problem when inserted into higher order graphs. Let us restate this problem here. Consider a graph with a line-like subgraph (\ref{fig:genertildeu}) 
on Moyal space,  where each of the $r+1$ little subgraphs is the  nonlocal tadpole (\ref{fig:tadpole}), 
\beq\label{fig:multTad}
\cdots \qquad 
\begin{picture}(130,20)(0,0)
\put(-10,-2){\line(0,1){5}} 
\put(0,2){\circle{2}}
\put(10,2){\circle{2}}
\put(20,2){\circle{2}}
\put(30,2){\circle{2}}
\put(13,-12){\small{$x_0$}}
\put(40,-2){\line(0,1){5}} 
\qbezier(10,2)(20,21)(30,2)
\qbezier(20,2)(35,28)(50,2)
\qbezier[5](-10,10)(-12,12)(-16,2)
\qbezier(-10,10)(-8,13)(0,2)
\put(50,2){\circle{2}}
\put(60,2){\circle{2}}
\put(70,2){\circle{2}}
\put(80,2){\circle{2}}
\put(63,-12){\small{$x_1$}}
\put(90,-2){\line(0,1){5}} 
\qbezier(60,2)(70,21)(80,2)
\qbezier(70,2)(85,28)(100,2)
\put(100,2){\circle{2}}
\put(110,2){\circle{2}}
\put(120,2){\circle{2}}
\put(130,2){\circle{2}}
\put(113,-12){\small{$x_2$}}
\put(140,-2){\line(0,1){5}} 
\qbezier(110,2)(120,21)(130,2)
\qbezier(120,2)(130,21)(140,10)
\qbezier[5](140,10)(142,9)(146,2)
\end{picture} 
\
\qquad \cdots 
\ 
\begin{picture}(150,20)(-20,0)
\put(-10,-2){\line(0,1){5}} 
\put(0,2){\circle{2}}
\put(10,2){\circle{2}}
\put(20,2){\circle{2}}
\put(30,2){\circle{2}}
\put(13,-12){\small{$x_{r-1}$}}
\put(40,-2){\line(0,1){5}} 
\qbezier[5](-10,10)(-12,12)(-16,2)
\qbezier(-10,10)(-8,13)(0,2)
\qbezier(10,2)(20,21)(30,2)
\qbezier(20,2)(35,28)(50,2)
\put(50,2){\circle{2}}
\put(60,2){\circle{2}}
\put(70,2){\circle{2}}
\put(80,2){\circle{2}}
\put(63,-12){\small{$x_{r}$}}
\put(90,-2){\line(0,1){5}} 
\qbezier(60,2)(70,21)(80,2)
\qbezier(70,2)(85,28)(100,2)
\put(100,2){\circle{2}}
\put(110,2){\circle{2}}
\put(120,2){\circle{2}}
\put(130,2){\circle{2}}
\put(113,-12){\small{$x_{r+1}$}}
\put(140,-2){\line(0,1){5}} 
\put(110,2){\line(0,1){5}}
\put(108.5,8){$\vdots$}
\put(120,2){\line(0,1){5}}
\put(118.5,8){$\vdots$}
\put(130,2){\line(0,1){5}}
\put(128.5,8){$\vdots$}
%
\end{picture} 
\qquad \cdots 
\eeq
The rest of the graph remains unspecified. Applying the rules from the appendix we find the, so far formal, expression  for this graph,
\beqan\label{eq:lineEuTad}
\int   w(x_0,x_{r+1};q_1,\dots,q_4)\ 
v (x_0,\dots,x_{r+1};q_1,\dots , q_4)
\qquad\qquad&& \nonumber\\
g(x_0) \cdots g(x_{r+1}) 
\ dx_0  \cdots dx_{r+1}\ dq_1\cdots dq_4
\quad &&
\eeqan
with $w$ denoting the distribution which corresponds to those parts of the graph that remain unspecified, and with $v$ given by
\beqan\label{eq:vlineTad}
&&\hspace{-2ex}v(x_0,\dots,x_{r+1};q_1,\dots , q_4) \ = 
\\ && \int 
\exp \big(-{\textstyle \frac \iu 2} q_1 \theta p_0 - \iu \sum_{j=0}^r  k_j \theta  p_{j} +{\textstyle \frac \iu 2} \sum_{j=0}^{r-1}  p_j \theta  p_{j+1} +{\textstyle \frac \iu 2} p_{r}\theta(q_2+q_3+q_4) \big) 
\ 
\nonumber \\ &&\quad \times \ \exp\big( - \iu  \sum_{j=0}^{r}   p_j(x_j-x_{j+1}) \, \big) \
\prod_{j=0}^{r} \; \frac 1 {p_j^2+m^2}\,   \frac 1 {k_j^2+m^2}  \
dp_0 \cdots dp_{r}\,dk_0 \cdots dk_{r}
\nonumber 
\eeqan 
Here, $k_j$ denotes the momentum corresponding to the edge that connects the two dots within vertex $x_j$, and $p_j$ the one of the edge connecting vertex $x_j$ with vertex $x_{j+1}$. The edge attached to the first dot at vertex $x_0$ has momentum $q_1$, and the edges that are attached to the three remaining dots of vertex $x_{r+1}$ are labelled by  $q_2,\dots,q_4$.
Observe that the only dependence on $q_1,\dots , q_4$ which we have put into $v$ is that of twisting factors involving also momenta $p_j$. In particular, the twisting that only involves $q_2,q_3,q_4$
will be contained in the distribution $w$.  To understand the notation, observe that  if, say, the as yet unspecified dot in vertex $x_0$ 
would correspond to an external leg with momentum $k$, then $w$ would contain the distribution $\delta^{(4)}(q_1-k)\, e^{-iq_1 x_0}$ and would be otherwise be independent of $x_0$ and $q_1$.

We will now understand $v$ in terms of distributions. In order to do so, we will use the $\delta$-distribution which we will eventually find in the adiabatic limit to simplify the twisting, as announced in  remark~\ref{rem:adiabTrick}.
In the distribution $v$ from (\ref{eq:vlineTad}) above, the adiabatic limit produces\footnote{In the usual language of Euclidean quantum field theory this corresponds to `calculating' the integrals $\int \exp\,\big(- \iu  \sum
 p_j(x_j-x_{j+1}) \big) \  dx_1 \cdots dx_r$}
$\delta$-distributions 
\[
\delta(p_j-p_{j-1}) \ , \qquad j=1, \dots , r
\]
such that the part of the twisting in $v$ which involves momenta from different vertices,
\[
\exp \big( +{\textstyle \frac \iu 2} \sum_{j=0}^{r-1}  p_j \theta  p_{j+1}  \big) 
\]
is 1 in the limit, by the antisymmetry of $\theta$. In the spirit of remark~\ref{rem:adiabTrick}, we discard this part of the twisting. 
Now, in the simplified expression for $v$, replace the formal integrals 
\[
\int\prod_j  e^{-ik_j \theta p_j} \, \frac 1 {k_j^2+m^2}\ dk_0 \cdots dk_r
\]
by the tensorproduct of distributions,
\[
G_E(\theta p_0) \cdots G_E(\theta p_r)
\]
We then find in the simplified expression for $v$, a product of distributions (the subscript $ET$ stands for Euclidean tadpole),
\beq\label{eq:produET}
 \prod_{j=1}^{r-1} u_{ET}(x_j-x_{j+1})
\eeq
where $u_{ET}$ is the (inverse) Fourier transform of the tempered distribution 
\[
\widetilde{u_{ET}} := \widetilde{G_E} \ \cdot \ \theta^* G_E\ , \mbox{ or formally, }
\widetilde{u_{ET}} (p) = \frac 1 {p^2+m^2} \ G_E (\theta p)  \ . 
\]
and two additional contributions $u_0(x_0-x_1)$ and $u_r(x_r-x_{r+1})$ which are the Fourier transforms of the tempered distributions
$
\widetilde{u_{ET}} \, e^{+\frac \iu 2 \langle \cdot , \theta q_1\rangle}$ 
and 
$\widetilde{u_{ET}} \, e^{+\frac \iu 2 \langle \cdot , \theta (q_2+q_3+q_4) \rangle}$, 
respectively.

Now, in the discussion of ordinary field theory, we have seen that in the adiabatic limit, a line-like graph produces products of the Fourier transforms of the distributions that correspond to the little subgraphs, cf.  (\ref{eq:line_moment}). 
This is still true in the present setting, so, in the adiabatic limit, the graph (\ref{fig:multTad}) produces products
of the distribution 
$
\widetilde{u_{ET}}
$. 
Now, $\widetilde{u_{ET}}$ contains the  {\em position space} propagator $G_E$ (or rather its pullback along $\theta$), and as we have already discussed in the previous section, for $d\geq 4$, products of $\theta^*G_E$ are only defined on $\dot \R^d$ and require renormalization when extended to all of $\R^d$.

It follows that, if  $p=0$ is in the domain of integration in  (\ref{eq:line_moment}), these products of distributions are ill-defined. It has been discussed elsewhere, e.g.~\cite{uvir}, that $p=0$ is in the domain of integration, for instance, in graphs of the form
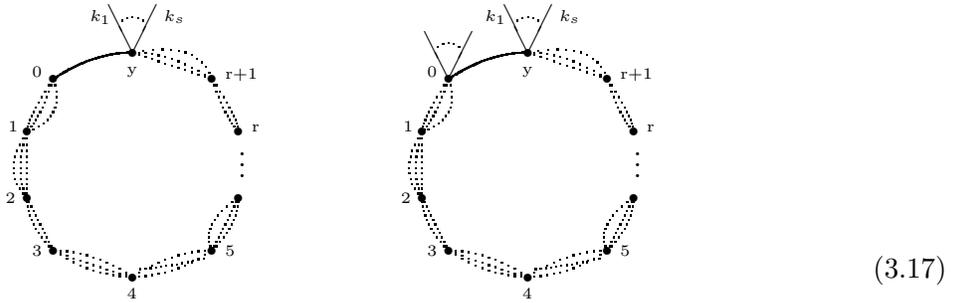
\beq\label{fig:circlegraph}
\begin{picture}(40,105)(0,0)
\put(30,10){\circle*{3}}
\put(-30,10){\circle*{3}}
\put(40,30){\circle*{3}}
\put(-40,30){\circle*{3}}
\put(40,55){\circle*{3}}
\put(-40,55){\circle*{3}}
\put(30,75){\circle*{3}}
\put(-30,75){\circle*{3}}
\put(0,85){\circle*{3}}
\put(0,0){\circle*{3}}
\qbezier[12](0,0)(14.5,0)(30,10)
\qbezier[12](0,0)(14.5,10)(30,10)
\qbezier[12](3,3)(14.5,4)(27,7)
\qbezier[12](0,0)(-14.5,0)(-30,10)
\qbezier[12](0,0)(-14.5,10)(-30,10)
\qbezier[12](0,0)(-14.5,4)(-30,10)
\qbezier[12](-30,10)(-35,20)(-40,30)
\qbezier[12](-30,10)(-40,20)(-40,30)
\qbezier[12](30,10)(25,20)(40,30)
\qbezier[12](30,10)(35,20)(40,30)
\qbezier[12](30,10)(40,20)(40,30)
\qbezier[12](-40,30)(-50,40)(-40,55)
\qbezier[12](-40,30)(-45,40)(-40,55)
\qbezier[12](-40,30)(-40,40)(-40,55)
\qbezier[12](-40,55)(-25,60)(-30,75)
\qbezier[12](-40,55)(-35,60)(-30,75)
\qbezier[12](-40,55)(-40,60)(-30,75)
\qbezier[12](40,55)(35,60)(30,75)
\qbezier[12](40,55)(40,60)(30,75)
\qbezier[12](0,85)(14.5,85)(30,75)
\qbezier[12](0,85)(25,90)(30,75)
\qbezier[12](0,85)(14.5,80)(30,75)
\put(0,85){\line(1,2){9}}
\put(0,85){\line(-1,2){9}}
\qbezier[4](-4,97)(0,100)(4,97)
\qbezier(-30,75)(-15,85)(0,85)
\put(-2,-8){\tiny{4}}
\put(35,8){\tiny{5}}
\put(-38,8){\tiny{3}}
\put(-48,28){\tiny{2}}
\put(40,38){$\vdots$}
\put(-2,77){\tiny{y}}
\put(35,75){\tiny{r+1}}
\put(-38,76){\tiny{0}}
\put(45,55){\tiny{r}}
\put(-47,55){\tiny{1}}
\put(-16,97){\tiny{$k_1$}}
\put(12,97){\tiny{$k_s$}}
\end{picture} 
\qquad \qquad \qquad \qquad \qquad 
\begin{picture}(40,105)(0,0)
\put(30,10){\circle*{3}}
\put(-30,10){\circle*{3}}
\put(40,30){\circle*{3}}
\put(-40,30){\circle*{3}}
\put(40,55){\circle*{3}}
\put(-40,55){\circle*{3}}
\put(30,75){\circle*{3}}
\put(-30,75){\circle*{3}}
\put(0,85){\circle*{3}}
\put(0,0){\circle*{3}}
\qbezier[12](0,0)(14.5,0)(30,10)
\qbezier[12](0,0)(14.5,10)(30,10)
\qbezier[12](3,3)(14.5,4)(27,7)
\qbezier[12](0,0)(-14.5,0)(-30,10)
\qbezier[12](0,0)(-14.5,10)(-30,10)
\qbezier[12](0,0)(-14.5,4)(-30,10)
\qbezier[12](-30,10)(-35,20)(-40,30)
\qbezier[12](-30,10)(-40,20)(-40,30)
\qbezier[12](30,10)(25,20)(40,30)
\qbezier[12](30,10)(35,20)(40,30)
\qbezier[12](30,10)(40,20)(40,30)
\qbezier[12](-40,30)(-50,40)(-40,55)
\qbezier[12](-40,30)(-45,40)(-40,55)
\qbezier[12](-40,30)(-40,40)(-40,55)
\qbezier[12](-40,55)(-25,60)(-30,75)
\qbezier[12](-40,55)(-35,60)(-30,75)
\qbezier[12](-40,55)(-40,60)(-30,75)
\qbezier[12](40,55)(35,60)(30,75)
\qbezier[12](40,55)(40,60)(30,75)
\qbezier[12](0,85)(14.5,85)(30,75)
\qbezier[12](0,85)(25,90)(30,75)
\qbezier[12](0,85)(14.5,80)(30,75)
\put(0,85){\line(1,2){9}}
\put(0,85){\line(-1,2){9}}
\qbezier[4](-4,97)(0,100)(4,97)
\put(-30,75){\line(1,2){9}}
\put(-30,75){\line(-1,2){9}}
\qbezier[4](-34,87)(-30,90)(-26,87)
\qbezier(-30,75)(-15,85)(0,85)
\put(-2,-8){\tiny{4}}
\put(35,8){\tiny{5}}
\put(-38,8){\tiny{3}}
\put(-48,28){\tiny{2}}
\put(40,38){$\vdots$}
\put(-2,77){\tiny{y}}
\put(35,75){\tiny{r+1}}
\put(-38,76){\tiny{0}}
\put(45,55){\tiny{r}}
\put(-47,55){\tiny{1}}
\put(-16,97){\tiny{$k_1$}}
\put(12,97){\tiny{$k_s$}}
\end{picture} 
\eeq
where the solid line represents the distribution $G_E(x_0-y)$.

Therefore, the insertion of two or more nonlocal tadpole graphs into the the line-like graph (\ref{fig:multTad}) turns out to be ill-defined in the adiabatic limit, if the line-like graph occurs in a graph of the form (\ref{fig:circlegraph}). It is the ultraviolet divergence of the position space propagator $G_E$, which now occurs at {\em momentum} $p=0$ in such graphs. This is why  one speaks of a mixing of ultraviolet and infrared divergences. 
It is impossible to renormalize this divergence locally, in the sense discussed in section 2, i.e. with counterterms given as linear combinations of derivatives of the $\delta$-distribution (in position space). Instead, counterterms of the form $\delta^{(l)}(p)$ or $\delta^{(l)}(\theta p)$ would be needed which correspond to position space expressions supported on all of $\R^d$.
%



Also note that arbitrarily large scaling degrees occur: an $r$-fold insertion of tadpole graphs leads to an $r$-fold product $\widetilde{u_{ET}}$ in the adiabatic limit, and hence to a distribution $\theta^*G_E^r$ which has scaling degree $r(d-2)$. So, the singularity in 0 becomes worse and worse, when more graphs are inserted -- which is especially peculiar, as the decrease at infinity in (\ref{eq:line_moment}) becomes better and better in that case.
On the other hand, the growth of the scaling degree (in 0) is not necessarily a problem in itself:  if it were only the tadpole which causes a problem, 
then it might be possible to consistently get rid of this tadpole from the perturbative expansion by a (nonlocal) redefinition of the product of fields. In such a theory, the divergence discussed above would no longer appear.

But the problem in fact occurs more generally. In the next section, we will briefly consider another example that has been discussed in the literature before. Again we will state the problem in terms of distributions and their wavefront sets.


\subsection{Nonlocal Fish}

Consider the following two fish graph contributions (in $\varphi^3$- and $\varphi^4$-theory), 
\beq\label{fig:twistFish}
\begin{picture}(100,20)(0,-5)
\put(10,2){\circle{2}}
\put(20,2){\circle{2}}
\put(30,2){\circle{2}}
\put(13,-12){\small{$x$}}
\put(40,-2){\line(0,1){5}} 
\qbezier(20,2)(35,28)(50,2)
\qbezier(30,2)(45,28)(60,2)
\put(50,2){\circle{2}}
\put(60,2){\circle{2}}
\put(70,2){\circle{2}}
\put(63,-12){\small{$y$}}
\end{picture} 
\mbox{ and } 
\begin{picture}(90,20)(-20,-5)
\put(0,2){\circle{2}}
\put(10,2){\circle{2}}
\put(20,2){\circle{2}}
\put(30,2){\circle{2}}
\put(13,-12){\small{$x$}}
\put(40,-2){\line(0,1){5}} 
\qbezier(20,2)(35,28)(50,2)
\qbezier(30,2)(45,28)(60,2)
\put(50,2){\circle{2}}
\put(60,2){\circle{2}}
\put(70,2){\circle{2}}
\put(80,2){\circle{2}}
\put(63,-12){\small{$y$}}
\end{picture} 
\eeq
We apply the rules and again simplify the twisting using the adiabatic limit as explained in remark \ref{rem:adiabTrick}. Let $k$ and $k^\prime$ label the external momenta in the graph from $\varphi^3$-theory, and for the graph from  $\varphi^4$-theory, let $k=k_1+k_2$, $k^\prime=k^\prime_1+k^\prime_2$ where $k_i$ and $k^\prime_i$ label the external momenta.
The formal expressions we then find for the above graphs are 
\[
\int e^{ikx} \, e^{ik^\prime y} \, u_{EF}(x-y) \, g(x) \, g(y) \ dx dy
\]
where 
\beq\label{eq:nonplFishEu}
u_{EF}(x) =  \int \widetilde{G_E} (p_1-p_2)\, \widetilde{G_E}(p_2)\, e^{-\iu p_1\theta p_2} \; e^{-\iu p_1 x} \ dp_1 dp_2  
\eeq
Here, the subscript $EF$ denotes 'Euclidean fish'. Let us again understand this in terms of distributions. First note that in the absence of the twisting,  instead of $u_{EF}$ we would have $G_E^2$, which, in dimensions $d\geq 4$, is only defined on $\dot \R^d$ and requires renormalization.
Contrary to this, with the twisting present, in any (even) dimension, the expression
\beq\label{eq:nonplFishMomEu}
\widetilde{u_{EF}}(p) := \int \widetilde{G_E} (p-k)\, \widetilde{G_E}(k)\, e^{-\iu p\theta k} \ dk
\eeq
is defined as an oscillatory integral, as was first proved in~\cite{doescher}.
As a distribution with respect to $p$ it is tempered, and its Fourier transform is the distribution $u_{EF}$ which appears in (\ref{eq:nonplFishEu}). 

Now, we line up fish graphs (with as little twisting as possible) in  a line-like graph as follows 
\beq\label{fig:fishlinephi3}
\cdots \qquad 
\begin{picture}(140,30)(0,-10)
\put(-10,-2){\line(0,1){5}} 
\put(0,2){\circle{2}}
\put(10,2){\circle{2}}
\put(20,2){\circle{2}}
\put(8,-12){\small{$x_0$}}
\put(30,-2){\line(0,1){5}} 
\qbezier(20,2)(30,20)(40,2)
\qbezier[5](-10,10)(-12,12)(-16,2)
\qbezier(-10,10)(-8,13)(0,2)
\qbezier[5](-10,18)(-12,20)(-16,8)
\qbezier(-10,18)(-8,20)(10,2)
\put(40,2){\circle{2}}
\put(50,2){\circle{2}}
\put(60,2){\circle{2}}
\put(48,-12){\small{$x_1$}}
\put(70,-2){\line(0,1){5}} 
\qbezier(50,2)(65,28)(80,2)
\qbezier(60,2)(75,28)(90,2)
\put(80,2){\circle{2}}
\put(90,2){\circle{2}}
\put(100,2){\circle{2}}
\put(110,-2){\line(0,1){5}} 
\put(88,-12){\small{$x_2$}}
\qbezier(100,2)(110,20)(120,2)
\put(120,2){\circle{2}}
\put(130,2){\circle{2}}
\put(140,2){\circle{2}}
\put(150,-2){\line(0,1){5}} 
\qbezier[5](150,10)(154,12)(156,2)
\qbezier(140,2)(148,13)(150,10)
\qbezier[5](150,10)(154,12)(160,4)
\qbezier(130,2)(138,20)(151,10)
\end{picture} 
\
\qquad \cdots 
\ 
\begin{picture}(140,30)(-20,-10)
\put(-10,-2){\line(0,1){5}} 
\put(0,2){\circle{2}}
\put(10,2){\circle{2}}
\put(20,2){\circle{2}}
\put(5,-12){\small{$x_{r-1}$}}
\put(30,-2){\line(0,1){5}} 
\qbezier(20,2)(30,20)(40,2)
\qbezier[5](-10,10)(-12,12)(-16,2)
\qbezier(-10,10)(-8,13)(0,2)
\qbezier[5](-10,10)(-12,12)(-20,4)
\qbezier(-10,10)(-2,20)(10,2)
\put(40,2){\circle{2}}
\put(50,2){\circle{2}}
\put(60,2){\circle{2}}
\put(48,-12){\small{$x_r$}}
\put(70,-2){\line(0,1){5}} 
\qbezier(50,2)(65,28)(80,2)
\qbezier(60,2)(75,28)(90,2)
\put(80,2){\circle{2}}
\put(90,2){\circle{2}}
\put(100,2){\circle{2}}
\put(88,-12){\small{$x_{r+1}$}}
\put(110,-2){\line(0,1){5}} 
\qbezier[5](110,10)(114,12)(116,2)
\qbezier(100,2)(108,13)(110,10)
%
\end{picture} 
 \cdots \quad 
\eeq
for $\varphi^3$ theory, and 
\beq\label{fig:fishlinephi4}
\cdots \qquad 
\begin{picture}(130,30)(0,-10)
\put(-10,-2){\line(0,1){5}} 
\put(0,2){\circle{2}}
\put(10,2){\circle{2}}
\put(20,2){\circle{2}}
\put(30,2){\circle{2}}
\put(13,-12){\small{$x$}}
\put(40,-2){\line(0,1){5}} 
\qbezier(20,2)(35,28)(50,2)
\qbezier(30,2)(45,28)(60,2)
\qbezier[5](-10,10)(-12,12)(-16,2)
\qbezier(-10,10)(-8,13)(0,2)
\qbezier[5](-10,18)(-12,20)(-16,8)
\qbezier(-10,18)(-8,20)(10,2)
%
\put(50,2){\circle{2}}
\put(60,2){\circle{2}}
\put(70,2){\circle{2}}
\put(80,2){\circle{2}}
\put(63,-12){\small{$x_1$}}
\put(90,-2){\line(0,1){5}} 
\qbezier(70,2)(85,28)(100,2)
\qbezier(80,2)(95,28)(110,2)
\put(100,2){\circle{2}}
\put(110,2){\circle{2}}
\put(120,2){\circle{2}}
\put(130,2){\circle{2}}
\put(113,-12){\small{$x_2$}}
\put(140,-2){\line(0,1){5}} 
%
%
\qbezier[5](140,10)(144,12)(146,2)
\qbezier(130,2)(138,13)(140,10)
\qbezier[5](140,10)(144,12)(150,4)
\qbezier(120,2)(128,20)(141,10)
\end{picture} 
\
\qquad \cdots 
\ 
\begin{picture}(150,30)(-20,-10)
\put(-10,-2){\line(0,1){5}} 
\put(0,2){\circle{2}}
\put(10,2){\circle{2}}
\put(20,2){\circle{2}}
\put(30,2){\circle{2}}
\put(13,-12){\small{$x_{r-1}$}}
\put(40,-2){\line(0,1){5}} 
\qbezier(20,2)(35,28)(50,2)
\qbezier(30,2)(45,28)(60,2)
\qbezier[5](-10,10)(-12,12)(-16,2)
\qbezier(-10,10)(-8,13)(0,2)
\qbezier[5](-10,10)(-12,12)(-20,4)
\qbezier(-10,10)(-2,20)(10,2)
\put(50,2){\circle{2}}
\put(60,2){\circle{2}}
\put(70,2){\circle{2}}
\put(80,2){\circle{2}}
\put(63,-12){\small{$x_r$}}
\put(90,-2){\line(0,1){5}} 
\qbezier(70,2)(85,28)(100,2)
\qbezier(80,2)(95,28)(110,2)
\put(100,2){\circle{2}}
\put(110,2){\circle{2}}
\put(120,2){\circle{2}}
\put(130,2){\circle{2}}
\put(113,-12){\small{$y$}}
\put(140,-2){\line(0,1){5}} 
\qbezier[5](140,10)(144,12)(146,2)
\qbezier(130,2)(138,13)(140,10)
\qbezier[5](140,18)(144,20)(146,8)
\qbezier(120,2)(138,20)(140,18)
%
\end{picture} 
\qquad \cdots 
\eeq
for $\varphi^4$, respectively. Applying again the trick of simplifying the twisting using the adiabatic limit, we find that also in this example, the subgraphs decouple, such that, with respect to $x_1, \dots, x_r$ the distribution corresponding to these graphs are
\[
\prod_{j=1}^{r-1} u_{EF}(x_j - x_{j+1})
\]
for the graph from $\varphi^4$-theory, and
\[
\prod_{j=1, {\rm odd} }^{r-1} u_{EF}(x_j - x_{j+1})\, \prod_{j=2, {\rm even} }^{r-1} G_{E}(x_j - x_{j+1})
\] 
for the graph from $\varphi^3$-theory.

In the adiabatic limit,  these line-like graphs will therefore produce products $\widetilde{u_{EF}}^k$ where $k=r+1$ 
in $\varphi^4$ theory, and $k=\left \lfloor \frac{r+1}{2} \right\rfloor$ in $\varphi^3$ theory. 

Now, observe that, as an oscillatory integral, the wavefront set of $\widetilde{u_{EF}}$ is contained in the manifold of stationary phase~\cite{hoerm},  so (for phase function $\phi(p,\xi)= p\theta \xi$)
\beqa
WF(\widetilde{u_{EF}}) &\subseteq &
\{\, (p; \nabla_p \phi(p,\xi) )\in \R^d \times \dot \R^d \ \big| \ \nabla_\xi\phi(p,\xi)=0 \, \} 
\\ & =&
\{\, (p; \theta \xi) \in \R^d \times \dot \R^d \ \big| \ \theta p=0 \, \} 
= \{\, (0;\xi) \in \R^d \times \dot \R^d \, \} 
\eeqa
where the last equality follows from the non-degenerateness of $\theta$. From (\ref{eq:nonplFishMomEu}) it is easily calculated that $\widetilde{u_{EF}}$'s scaling degree in 0 is $d-4$  in $d$ dimensions.

It follows that in $d=4$ dimensions, monomials of $\widetilde{u_{EF}}$ of arbitrarily high order all have the same scaling degree $0$ and can be extended uniquely. For  $d\geq 6$, on the other hand, the $k$-fold product has scaling degree $k(d-4)>0$
and indeed needs renormalization for $k \geq d/(d-4)$ (e.g. $k \geq 3$ in $d=6$). Note that the scaling degree grows to arbitrarily high orders in this case.  For a related discussion in $d=4$ and $6$, see~\cite{doescher}.


Again, it is emphasized that the growth of the scaling degree in itself does not necessarily mean that all is lost: If one admits nonlocal counterterms, there might be a nonlocal subtraction which gives a modified (still ultraviolet-regular) nonlocal fish graph, whose Fourier transform is well-behaved.  In a way, this is what was achieved in ~\cite{GrW,R} for all graphs, by means of a modified propagator which compensates all mixing divergences.


\subsection{Minkowski signature}

We now turn to theories with hyperbolic signature (Minkowskian Moyal space), and study the insertion of the 
tadpole and fish graph corresponding to the examples we studied in the Euclidean setting.
For the tadpole insertion, it is proved that there is no mixing problem, and for the fish graph, some evidence is given that also here,  the mixing might be absent.

For the single nonplanar tadpole
\[
\begin{picture}(30,10)
\put(0,2){\circle{2}}
\put(10,2){\circle{2}}
\put(20,2){\circle{2}}
\put(30,2){\circle{2}}
\qbezier(10,2)(20,21)(30,2)
\end{picture} 
\]
there is no time ordering at the same vertex (only different vertices can be related to each other as being later or earlier), so instead of a propagator as in (\ref{eq:tadpoleEuNoAL}), we now find the formal expression
\beqan\label{eq:tadpoleMi}
\int \frac 1 {2 \omega_{\ra p}} \ e^{- \iu (q^\prime+q)x} \ 
e^{-\frac \iu 2 q^\prime \theta q- \iu  \tilde p \theta q }
\ g(x)  \ d \ra p\, dx 
& =& 
\tilde g (q^\prime+q) e^{-\frac \iu 2 q^\prime \theta q} \ \Delta_+(\theta q)
\eeqan
with the (not-time-ordered) 2-point function  
\beq\label{eq:Dp_OsciInt}
\Delta_+(x)= \int \frac 1 {2 \omega_{\ra p}} \  e^{- \iu \tilde p x } \; d \ra p \qquad \mbox{ with } 
\tilde p=(\omega_{\ra p}, \ra p) \in \R_{>0}\times \R^{d-1}
\eeq
given as an oscillatory integral (cf. \cite[Chap X]{RS}). As such, its  wavefront set is contained in its manifold of stationary phase,
\beqan\label{eq:WF_Delta+}
WF(\Delta_+) &\subseteq &\{ (0,0;|\ra p|, \ra p) \in  \R^d \times \dot \R^d \; | \; \ra p \neq 0 \}
\nonumber 
\\&& \cup \{ (\pm |\ra x|, \ra x; \lambda |\ra x|, \mp \lambda \ra x) \in \R^d \times \dot \R^d \; | \; \lambda>0, \ra x \neq 0 \}
\eeqan
The notation is redundant in the sense that if  $(|\ra p|, \ra p)$ and $(\lambda |\ra x|, \mp \lambda \ra x)  \in \dot \R^d$, then clearly $\ra p$ and $\ra x$ must be non-zero, a fact which, however, we would like to emphasize. 
Observe that $|\ra p|$ and $\lambda |\ra x|$ are (strictly) positive in the above, so from H\"ormander's criterion it follows that in arbitrary dimension $d$, products of $\Delta_+$ are well-defined distributions on $\R^d$. By our general argument regarding the wavefront sets of pullbacks along non-degenerate linear maps (\ref{eq:WF_Au}), this is also true for products of $\theta^*\Delta_+$.

Let us now line up such graphs as we have done it in the Euclidean framwork (\ref{fig:multTad}), 
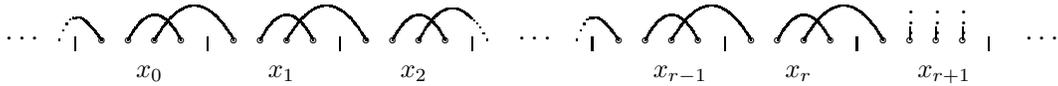
\[
\cdots \qquad 
\begin{picture}(130,20)(0,0)
\put(-10,-2){\line(0,1){5}} 
\put(0,2){\circle{2}}
\put(10,2){\circle{2}}
\put(20,2){\circle{2}}
\put(30,2){\circle{2}}
\put(13,-12){\small{$x_0$}}
\put(40,-2){\line(0,1){5}} 
\qbezier(10,2)(20,21)(30,2)
\qbezier(20,2)(35,28)(50,2)
\qbezier[5](-10,10)(-12,12)(-16,2)
\qbezier(-10,10)(-8,13)(0,2)
\put(50,2){\circle{2}}
\put(60,2){\circle{2}}
\put(70,2){\circle{2}}
\put(80,2){\circle{2}}
\put(63,-12){\small{$x_1$}}
\put(90,-2){\line(0,1){5}} 
\qbezier(60,2)(70,21)(80,2)
\qbezier(70,2)(85,28)(100,2)
\put(100,2){\circle{2}}
\put(110,2){\circle{2}}
\put(120,2){\circle{2}}
\put(130,2){\circle{2}}
\put(113,-12){\small{$x_2$}}
\put(140,-2){\line(0,1){5}} 
\qbezier(110,2)(120,21)(130,2)
\qbezier(120,2)(130,21)(140,10)
\qbezier[5](140,10)(142,9)(146,2)
\end{picture} 
\
\qquad \cdots 
\ 
\begin{picture}(150,20)(-20,0)
\put(-10,-2){\line(0,1){5}} 
\put(0,2){\circle{2}}
\put(10,2){\circle{2}}
\put(20,2){\circle{2}}
\put(30,2){\circle{2}}
\put(13,-12){\small{$x_{r-1}$}}
\put(40,-2){\line(0,1){5}} 
\qbezier[5](-10,10)(-12,12)(-16,2)
\qbezier(-10,10)(-8,13)(0,2)
\qbezier(10,2)(20,21)(30,2)
\qbezier(20,2)(35,28)(50,2)
\put(50,2){\circle{2}}
\put(60,2){\circle{2}}
\put(70,2){\circle{2}}
\put(80,2){\circle{2}}
\put(63,-12){\small{$x_{r}$}}
\put(90,-2){\line(0,1){5}} 
\qbezier(60,2)(70,21)(80,2)
\qbezier(70,2)(85,28)(100,2)
\put(100,2){\circle{2}}
\put(110,2){\circle{2}}
\put(120,2){\circle{2}}
\put(130,2){\circle{2}}
\put(113,-12){\small{$x_{r+1}$}}
\put(140,-2){\line(0,1){5}} 
\put(110,2){\line(0,1){5}}
\put(109,4){$\vdots$}
\put(120,2){\line(0,1){5}}
\put(119,4){$\vdots$}
\put(130,2){\line(0,1){5}}
\put(129,4){$\vdots$}
%
\end{picture} 
\qquad \cdots 
\]

For the time ordering, implemented by Heaviside functions $\tau$, we choose  $x_{j,0}>x_{j+1,0}$ ($j=0,\dots,r$).
As in the Euclidean setting, we label the momentum that corresponds to the edge connecting vertex $x_j$ with vertex $x_{j+1}$ by the letter $p$ and an index $j$, and the momentum that corresponds to an edge within vertex $x_j$ by the letter $k$ and an index $j$. Observe that with our conventions, in the Minkowskian setting, all such momenta will now be on the positive mass shell, $\tilde p_j=(\omega_{\ra p_j}, \ra p_j )$, where $\ra p_j \in \R^{d-1}$ and $\omega_{\ra p_j}= \sqrt{\ra p_j^2 + m^2}$. We denote by $q_1$ and $q_2, q_3 , q_4$, the momenta of the edges attached to the first dot in  $x_0$ and the last three dots in $x_{r+1}$, respectively. Putting everything we left unspecified into a distribution $w$, we then find 
\beqan\label{eq:lineMinkTad}
\int  \prod_{j=0}^{r}\tau(x_{j,0}-x_{j+1,0}) \  w(x_0,x_{r+1};q_1,\dots,q_4)\ 
v (x_0,\dots,x_{r+1};q_1,\dots , q_4)\qquad\qquad&&
\nonumber\\
g(x_0) \cdots g(x_{r+1}) 
\ dx_0  \cdots dx_{r+1}\ dq_1\cdots dq_4
\quad &&
\eeqan
with the, for now formal, expression 
\beqa
&&\hspace{-2ex}v(x_0,\dots,x_{r+1};q_1,\dots , q_4) \ = 
\\ && \int 
\exp \big(-{\textstyle \frac \iu 2} q_1 \theta \tilde p_0  - \iu \sum_{j=0}^r  \tilde k_j \theta  \tilde p_{j} +{\textstyle \frac \iu 2} \sum_{j=0}^{r-1} \tilde p_j \theta \tilde p_{j+1} +{\textstyle \frac \iu 2} 
\tilde p_{r}\theta(q_2+q_3+q_4) \big) \ 
\\ &&\quad \times 
\exp\big(
 - \iu  \sum_{j=0}^{r}  \tilde p_j(x_j-x_{j+1})
  \, \big) 
\  \prod_{j=0}^{r} \frac 1 {2 \omega_{\ra p_j}}\ \frac 1 {2 \omega_{\ra h_j}} 
\ d \ra p_0  \cdots d \ra p_{r}  \  d \ra k_0  \cdots d \ra k_{r} 
\eeqa
Observe that -- as before in the Euclidean setting -- the only dependence on $q_1,\dots , q_4$ which was put into $v$ is that of twisting factors involving also momenta $\tilde p_j$.

We now simplify the twisting by means of the $\delta$-distri\-bu\-tions which appear in the adiabatic limit. In the Minkowskian setting, however, due to the fact that the time-ordering is separate, we find $\delta$-distributions only for the spatial parts: only 3-momenta are conserved at the vertex, although -- of course, by the theory's translation invariance -- the overall $4$-momentum is conserved. Explicitly, for the graph above, we have in the adiabatic limit,
\[
\delta^{(3)} (\ra p_1 - \ra p_{0}) \ \delta^{(3)} (\ra p_2 - \ra p_{1}) \cdots 
\delta^{(3)} (\ra p_{r+1} - \ra p_{r})
\]
It follows that the products $\tilde p_j \theta \tilde p_{j+1}$ for $j=0,\dots, r$ are in fact equal to 0 in the adiabatic limit. Note that contrary to the Euclidean situation, signs are very important here: a $\delta$-distribution of the form $\delta^{(3)} (\ra p + \ra p^\prime)$, does not make $\tilde p \theta \tilde p^\prime$ equal to 0,  since $\tilde p \theta (\omega_{-\ra p}, -\ra p) =  - 2 \omega_{\ra p} (\theta p)^0$. 

Now, in the simplified expression for $v$, replace the formal integrals regarding $\ra k_j$ by the following  tensorproduct of distributions,
\[
\Delta_+(\theta \tilde p_0)\,  \Delta_+(\theta \tilde p_1) \cdots  \Delta_+(\theta \tilde p_{r-1} ) \Delta_+(\theta \tilde p_{r})  
\]
We then find in the simplified expression for $v$, a product 
\beqa
\qquad  \prod_{j=1}^{r-1} v_T (x_j-x_{j+1}) \ 
\eeqa
where 
\[
v_T(x)= \int 
\frac 1 {2 \omega_{\ra p}}\  \Delta_+(\theta \tilde p)\; e^{-i\tilde px}\ d\ra p
\]
and, as in the Euclidean setting, two more distributions in $x_0-x_1$ and $x_r-x_{r+1}$, respectively, which depend on the momenta $q_1, \dots , q_4$. Now, since $\tilde p$ is on the mass-shell, $\theta \tilde p$ is spacelike (if $\theta$ is the standard symplectic matrix, we have $(\theta \tilde p)^2= -
m^2-p_1^2-p_3^2$), so $\Delta_+(\theta \tilde p)$ is actually a smooth, quickly decreasing  function in $(\theta \tilde p)^2$. Therefore, $v_T$ is defined as an oscillatory integral, and since $v_T$ without the smooth function $\theta^* \Delta_+(\tilde p)$ is the 2-point function $\Delta_+$, the wavefront set of $v_T$ is contained in  that of $\Delta_+$. Therefore, the product of $v_T$ with the Heaviside function is defined, and we find that in the full expression (\ref{eq:lineMinkTad}),  among other contributions, also a product of distributions
\[
\prod_{j=1}^{r-1} \tau(x_{j,0}-x_{j+1,0})\, v_T(x_{j,0}-x_{j+1,0})
\]  
occurs. 
The Fourier transform of $\tau(x_0)v_T(x)$ is
\[
\frac1 {\omega_{\ra p} -p_0 + \iu \epsilon} \ \frac 1 {2 \omega_{\ra p}} \ \Delta_+(\theta \tilde p) 
\]
Contrary to the Euclidean situation,  products of this distribution in $p$ are defined as distributions on all of $\R^4$.
This means that although the presence of $\Delta_+(\theta \tilde p)$ considerably modifies the theory's behaviour in the infrared regime~\cite{qplan1}, there is no mixing effect which would destroy renormalizability.

One might argue that the tadpole insertion is special in the sense that  no time ordering  is involved in the tadpole's inner edge and the 2-point function $\Delta_+$ is an especially harmless distribution. There is, however, some indication that also multiple insertions of those fish graphs which correspond to the Euclidean graphs we studied above, do not cause an infrared problem in the Hamiltonian setting. To see this, consider again the graphs (\ref{fig:fishlinephi3}) and (\ref{fig:fishlinephi4}) 
for $\varphi^3$- and $\varphi^4$-theory. We first discuss the graph from  $\varphi^3$-theory. For the time ordering we again choose the one where $x_{j,0}>x_{j+1,0}$. 
We label 
the single momentum leaving a vertex $x_j$, $1\leq j \leq r$ even, towards the right with an index $j$, and write $\tilde p^1_j$ and $\tilde p_j^2$ for the two momenta leaving a vertex $x_j$, $1\leq j \leq r$ odd, towards the right. From the rules we find the following twisting,
\[
\prod_{j=1, {\rm odd}}^r\exp\big(-{\textstyle \frac \iu 2} (-\tilde p_{j-1}+\tilde p_{j+1})\theta(\tilde p_j^1+\tilde p_j^2) - \iu \, \tilde p_j^1\theta \tilde p_j^2 \, \big)
\]
We again use the $\delta$-distributions $\delta(-\ra p_{j-1}+\ra p_j^1 + \ra p_j^2)$ that will occur in the adiabatic limit to simplify the twisting to
\[
\prod_{j=1, {\rm odd}}^r\exp\big( - \iu \, \tilde p_j^1\theta \tilde p_j^2 \, \big)
\]
As in the Euclidean case, we then find products of distributions $v_F$, 
\[
\prod v_F (x_j-x_{j+1})
\]
where, however, in the present setting, we have
\[
v_F(x) = \int e^{-\iu (\tilde p_1 + \tilde p_2) x}\,e^{-\iu \tilde p_1 \theta \tilde p_2}\, \frac 1 {2 \omega_{\ra p_1}} \, \frac 1 {2 \omega_{\ra p_2}} \ d\ra p_1 d \ra p_2
\]
This has been shown to be a tempered distribution in~\cite{bahns_schwinger}. 
In~\cite{schulz_BA}, for $\theta$ the standard symplectic matrix, and $d=4$, 
its wavefront set 
was shwon to be a subset of 
\[
WF(v_F) \subseteq WF(\Delta_+) \cup \{(x,p) \in \R^4\times \dot \R^4\ | \ |x_0| > |x_2|\, , \ p_0 \geq |\ra p| \, \}  
\]
The proof can be generalized in a straightforward way to higher (even) dimensions $d\geq 6$ with nondegenerate $\theta$ and  $\theta^{0j}=-\theta^{j0}= \delta_{j,2}$, and we find the wavefront set above, simply with $4$ replaced by $d$. 
By H\"ormander's criterion, the product with a Heaviside function,
\[
u_F(x):= \tau(x_0) \, v_F(x)
\]
is therefore a distribution on $\R^d$. 

Now, contrary  to the Euclidean Moyal counterpart (\ref{eq:nonplFishEu}), also its Fourier transform
\[
\widetilde{u_F}(p)= \int \frac 1 {p_0 - \omega_{\ra k} -\omega_{\ra p - \ra k}+\iu \epsilon}\, \frac 1  {2 \omega_{\ra p - \ra k}}
\, \frac 1  {2 \omega_{ \ra k}}\ e^{-\iu \tilde k \theta (\omega_{\ra p -\ra k }, \ra p - \ra k)}\ d\ra k
\]
seems to be well-behaved in $0$, independently of the dimension $d$, since, as we observed before,  $\tilde k \theta (\omega_{ -\ra k }, - \ra k) = -2\omega_{\ra k}(\theta k)_0$, so a part of the twisting factor remains even for $p=0$. Whether the oscillating factor really suffices to give this integral meaning as an oscillatory integral, is not clear -- in fact, essentially nothing seems to be known about oscillatory integrals with such complicated phases. This question will be addressed in a broader context in a future publication on properties of the phases which can appear in hyperbolic field theory on Moyal space\footnote{Work in progress joint with J. Zahn}.

Here, we only mention a formal calculation of $\widetilde{u_F}(0)$ in $d=4$ and $6$: using polar coordinates with 3-axis given by the vector $(\theta^{0i}) \in \R^{d-1}$, and splitting the radial integral into two parts one of which does not include 0, we find 
\[
\int \frac 1 {- 2\omega_{\ra k} }\, \frac 1  {4 \omega_{\ra k}^2}
\ e^{2\iu \omega_{\ra k}(\theta k)_0}\ d\ra k
\ = \  c + c^\prime\int_a^\infty \frac 1 {r^5 } \ r^{d-2}\ \ \sin(2r\,\sqrt{r^2+m^2}\,
)\; dr
\]
Observe that the additional factor $r^{-2}$ in the integrand on the right hand side 
is produced by the azimuthal angle integration. In $d=4$, the  integrand on the right hand side is now even integrable,
and in $d=6$, after a change of variables, we find the oscillatory integral $\int \frac{\sin(r)}{r}\, dr$.

The $\varphi^4$-graph is similar, although here, a remnant of the twisting which links different $v_F$'s remains even when we use the adiabatic limit to simplify it, such that 
in the adiabatic limit we find products of $\widetilde{u_F}(p)$ along with some remnant of the twisting, e.g. a power of $\exp(-2\iu \omega_{\ra p } \theta^{0j}p_j)$.


\section{A different mechanism}\label{sec:UVIRMink}

We will now see that a mixing problem of a different nature does occur in the Minkowskian setting, and that this effect is not present in the Euclidean setting.

Consider the following graphs
\begin{equation}\label{gr:single_phi3}
\begin{picture}(50,10)
\put(10,0){\circle*{2}}
\put(20,0){\circle*{2}}
\put(30,0){\circle*{2}}
\put(35,-3){\line(0,1){2}}
\put(40,0){\circle*{2}}
\put(50,0){\circle*{2}}
\put(60,0){\circle*{2}}
\qbezier(10,0)(30,23)(50,0)
\qbezier(20,0)(30,13)(40,0)
\end{picture}
\qquad \mbox{ and } \qquad 
\begin{picture}(50,10)
\put(10,0){\circle*{2}}
\put(20,0){\circle*{2}}
\put(30,0){\circle*{2}}
\put(40,0){\circle*{2}}
\put(45,-3){\line(0,1){2}}
\put(50,0){\circle*{2}}
\put(60,0){\circle*{2}}
\put(70,0){\circle*{2}}
\put(80,0){\circle*{2}}
\qbezier(10,0)(40,27)(70,0)
\qbezier(20,0)(40,20)(60,0)
\qbezier(30,0)(40,12)(50,0)
\end{picture}
\end{equation}
from $\varphi^3$, and in $\varphi^4$-theory, respectively. 
In the Euclidean setting on Moyal space, the simplified twisting of these graphs is 1, so that they are simply the ordinary fish and setting sun graph 
\[
\begin{picture}(130,30)(-30,0)
\put(0,20){\line(1,0){20}}
\put(20,20){\circle*{3}}
\put(40,20){\circle{140}}
\put(60,20){\circle*{3}}
\put(60,20){\line(1,0){20}}
\end{picture}
\begin{picture}(30,20)
\put(10,15){$\mbox{and}$} 
\end{picture}
\begin{picture}(130,30)(-35,0)
\put(0,20){\line(1,0){80}}
\put(20,20){\circle*{3}}
\put(40,20){\circle{140}}
\put(60,20){\circle*{3}}
\end{picture}
\]
and as such would require ordinary local renormalization.  In the Minkowskian regime, on the other hand, a remnant of the twisting remains even in the adiabatic limit. We shall see now that it renders both graphs finite (in any dimension $d$).
Consider first the twistings in the two respective graphs,
\[
e^{-\frac \iu 2 (\tilde p_1 + \tilde p_2) \theta (k-k^\prime)} 
\qquad \mbox{ and } \qquad 
e^{-\frac \iu 2 (\tilde p_1 + \tilde p_2 + \tilde p_3) \theta (k-k^\prime)} 
\]
where $k$ and $k^\prime$  label the external momenta. Using again the $\delta$-distributions from the adiabatic limit, they can be simplified to
\[
e^{- \iu  (\omega_{\ra p_1}+\omega_{\ra p_2}) \theta^{0j}k_j}   \; 
e^{- \iu  k_{0} \theta^{0j}k_j}  
\qquad \mbox{ and } \qquad 
e^{- \iu  (\omega_{\ra p_1}+\omega_{\ra p_2}+\omega_{\ra p_3}) \theta^{0j}k_j}   \; 
e^{- \iu  k_{0} \theta^{0j}k_j}  
\]
So, we find the following, for now formal, expressions
\beq\label{eq:plFM}
\int e^{-\iu k x} \, e^{-\iu k^\prime y}\; 
e^{- \iu  k_{0} \theta^{0j}k_j} \  \tau(x_0-y_0) \ \Delta^s_+(x_0-y_0 + (\theta k)_0, \ra x - \ra y)\;  g(x)\, g(y) \ dx dy
\eeq
with $s=2$ for the first and $s=3$ for the second graph. 
To give meaning to the above, we will treat it as a distribution with respect to {\em both} position space and the external momentum, or rather in $(\theta k)_0$. To make things easier, we choose $\theta$ to be the standard symplectic matrix, so we have $(\theta k)_0=
k_2$. It makes sense to treat the expression as a distribution also in momentum space, since in the full theory, for external legs, one would actually consider (quasiplanar) Wick products instead of simple exponentials. These operator-valued distributions would act on a suitable domain in Fock space, thereby producing additional functions (wavefunctions) in the external momenta as well as corresponding integrations.

So,  consider the linear map $S: \R^{d}\times \R\rightarrow \R^d$, $S(x,a)=(x_0+ a,\bf x)$. Its set of normals is easily calculated to be $\{(x;0)\} \subseteq \R^{d}\times \R^{d}$, so as discussed  on page~\pageref{eq:WF_Au}, we can pull back any distribution on $\R^d$ along $S$. The resulting distribution's wavefront set is contained in (cf. to equation (\ref{eq:WF_Au}))
\beqa
WF(S^*u)\subseteq S^*(WF(u)) &=& 
\{(x,a;p,p_0) \in \R^{d+1} \times \dot \R^{d+1}\ | \ (x_0+a,\ra x; p) \in WF(u)\}
\\ &=& 
\{(x,a-x_0;p,p_0) \in \R^{d+1} \times \dot \R^{d+1}\ | \ (a,\ra x; p) \in WF(u)\}
\eeqa

For later purposes, we note that  we can  also pull back any distribution on $\R^d$ along  the linear map $T: \R^{d}\times \R\rightarrow \R^d$, $T(x,a)=(x_0- a,\bf x)$, and have
\beqa
T^*(WF(u)) &=& 
\{(x,x_0-a;p,-p_0) \in \R^{d+1} \times \dot \R^{d+1}\ | \ (a,\ra x; p) \in WF(u)\}
\eeqa

We now prove that the product of the Heaviside function with  $S^* (\Delta_+^s)$  and its product  with  $T^* (\Delta_+^s)$  is well-defined -- where $\tau$ is of course understood as a distribution on $\R^{d+1}$,
\[
\tau(g)=\int \tau(x_0) g(x_0,\ra x,a) \ dx da
\]
In terms of formal integral kernels, this means that $\tau(x_0) t_{-a} \Delta_+^s (x_0,\ra x)$ where $t_a$ denotes the translation by $a \in \R$ with respect to the first argument only, $t_a u(x)= u(x_0-a,\ra x)$, is a distribution on $\R^d \times \R$. As such, (\ref{eq:plFM}) makes sense. 

{\prop\label{prop:well} \rm{Let $\Delta_+$ denote the 2-point function on $\R^d$ and let $\tau$ denote the Heaviside function on $\R^{d+1}$ in the sense above. For $T: \R^{d}\times \R\rightarrow \R^d$, $T(x,a)=(x_0- a,\bf x)$ and 
$S: \R^{d}\times \R\rightarrow \R^d$, $S(x,a)=(x_0+ a,\bf x)$, we have 
$
\tau S^*\Delta_+ \in \mc D^\prime(\R^{d+1})$ and 
$\tau T^*\Delta_+ \in \mc D^\prime(\R^{d+1})$.
}}

\proof{
We first observe that, by (\ref{eq:WFprod}) and (\ref{eq:WF_Delta+}),  both $WF(\Delta_+^2)$ and $WF(\Delta_+^3)$ are contained in 
\beqa
&& \{ (0,0;p_0, \ra p) \in \R^d \times \dot \R^d \; | \; p_0> |\ra p|\geq 0 \} 
\ \cup \
\{ (\pm |\ra x|, \ra x; + \lambda|\ra x|,\mp \lambda \ra x ) \in \R^d \times \dot \R^d \; | \; \lambda >0, \ra x \neq 0\}
\eeqa
so for $s=2,3$,
\beqa
WF(S^*\Delta_+^s) &\subseteq &\{ (x_0,0,-x_0;p_0, \ra p ,p_0) \; | \; p_0 > |\ra p| \geq 0 \}
\\ && 
\cup \ 
\{ (x_0, \ra x, \pm |\ra x| -x_0\; ; \lambda |\ra x|, \mp \lambda \ra x, \lambda |\ra x|) \; | \; \lambda >0, \ra x \neq 0\}
\ \subset T^*(\R\times \R^{d} \times \R  )
\eeqa
and
\beqa
WF(T^*\Delta_+^s) &\subseteq& \{ (x_0,0,x_0;p_0, \ra p ,-p_0) \; | \; p_0 > |\ra p| \}
\\ && 
\cup \ 
\{ (x_0, \ra x, x_0\mp |\ra x| \; ; \lambda |\ra x|, \mp \lambda \ra x, -\lambda |\ra x|) \; | \; \lambda >0, \ra x \neq 0\}
\ \subset T^*(\R\times \R^{d} \times \R  )
\eeqa
Now, the wavefront set of the Heaviside function (as a distribution on $\R^{d+1}$) is 
\[
WF(\tau) = \{ (0,\ra x, a; \lambda,\ra 0, 0) \in \R^{d+1} \times \dot \R^{d+1} \ | \ \lambda \neq 0\}
\]
so, the intersection of the singular supports of $\tau$ and $S^*(\Delta_+^s)$ or $T^*(\Delta_+^s)$, $s=2,3$, respectively, is $\{(0,\ra x, \pm|\ra x|)\}$ in both cases, including both times also 
$\ra x =0$. Writing the corresponding momenta from the respective wavefront sets of $\tau$, $S^*(\Delta_+^s)$ and $T^*(\Delta_+^s)$ in a table, we find for $\ra x=0$,
\[
\begin{array}{r|ccc}
& \R & \R^{d-1} & \R 
\\\hline 
\tau &  \lambda \neq 0 & 0 & 0
\\
\hline
S^*\Delta_+^s & p_0 & \ra p & p_0
\\
&(p_0 > |\ra p| \geq 0)& 
\\
\hline T^*\Delta_+^s & k_0 & \ra k & -k_0
\\ 
& (k_0 > |\ra k| \geq 0)
\end{array}
\]
and for $\ra x\neq0$,
\[
\begin{array}{r|ccc}
& \R & \R^{d-1} & \R 
\\\hline 
\tau &  \lambda \neq 0 & 0 & 0
\\
\hline
S^*\Delta_+^s & \mu |\ra x| & \mp \mu \ra x & \mu| \ra x|
\\
&(\mu > 0)& 
\\
\hline T^*\Delta_+^s & \nu |\ra x| & \mp\nu \ra x & -\nu |\ra x|
\\ 
& (\nu > 0)
\end{array}
\]

We can read off that the momenta from $WF(\tau)$ cannot add up to 0 with those of $S^*(\Delta_+^s)$ or $T^*(\Delta_+^s)$, respectively. By H\"ormander's criterion, the claim follows. 

Observe that it is the translation by $a$ which makes this work -- if the last entry were not present, the momenta could indeed add up to 0, and we would find the ordinary fish and setting sun graph divergences. }

From the proposition, we deduce that the graphs (\ref{gr:single_phi3}) are ultraviolet-regular -- contrary to the ordinary fish and setting sun graphs.

By having considered not only the translation $S$, but also $T$ in the above proposition, we also conclude that the two graphs
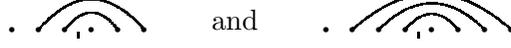
\[
\begin{picture}(50,10)
\put(0,0){\circle*{2}}
\put(10,0){\circle*{2}}
\put(20,0){\circle*{2}}
\put(25,-3){\line(0,1){2}}
\put(30,0){\circle*{2}}
\put(40,0){\circle*{2}}
\put(50,0){\circle*{2}}
\qbezier(10,0)(30,23)(50,0)
\qbezier(20,0)(30,13)(40,0)
\end{picture}
\qquad \mbox{ and } \qquad 
\begin{picture}(50,10)
\put(0,0){\circle*{2}}
\put(10,0){\circle*{2}}
\put(20,0){\circle*{2}}
\put(30,0){\circle*{2}}
\put(35,-3){\line(0,1){2}}
\put(40,0){\circle*{2}}
\put(50,0){\circle*{2}}
\put(60,0){\circle*{2}}
\put(70,0){\circle*{2}}
\qbezier(10,0)(40,27)(70,0)
\qbezier(20,0)(40,20)(60,0)
\qbezier(30,0)(40,12)(50,0)
\end{picture}
\]
from $\varphi^3$, and in $\varphi^4$-theory, respectively, are ultraviolet-regular, since their (simplified) analytic expressions are 
\[
\int e^{-\iu k x} \, e^{-\iu k^\prime y}\; 
e^{-2 \iu  k_{0} \theta^{0j}k_j} \  \tau(x_0-y_0) \ \Delta^s_+(x_0-y_0 - (\theta k)_0, \ra x - \ra y)\;  g(x)\, g(y) \ dx dy
\]
with $s=2$ for first and $s=3$ for the second graph.

However, as we have seen in the proof of the proposition, the signs in the wavefront sets of $T^*\Delta^s_+$ and $S^*\Delta^s_+$ differ. This will turn out to be problem in the following two graphs,
\begin{equation}\label{gr:ex_phi3}
\begin{picture}(120,20)
\put(10,0){\circle*{2}}
\put(20,0){\circle*{2}}
\put(30,0){\circle*{2}}
\put(35,-3){\line(0,1){2}}
\put(40,0){\circle*{2}}
\put(50,0){\circle*{2}}
\put(60,0){\circle*{2}}
\put(65,-3){\line(0,1){2}}
\put(70,0){\circle*{2}}
\put(80,0){\circle*{2}}
\put(90,0){\circle*{2}}
\put(95,-3){\line(0,1){2}}
\put(100,0){\circle*{2}}
\put(110,0){\circle*{2}}
\put(120,0){\circle*{2}}
\qbezier(10,0)(30,23)(50,0)
\qbezier(20,0)(30,13)(40,0)
\qbezier(60,0)(65,13)(70,0)
\qbezier(80,0)(100,23)(120,0)
\qbezier(90,0)(100,13)(110,0)
\end{picture}
\end{equation}
in $\varphi^3$ theory, and 
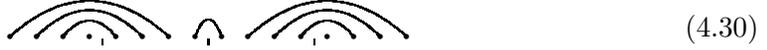
\begin{equation}\label{gr:ex_phi4}
\begin{picture}(160,30)
\put(10,0){\circle*{2}}
\put(20,0){\circle*{2}}
\put(30,0){\circle*{2}}
\put(40,0){\circle*{2}}
\put(45,-3){\line(0,1){2}}
\put(50,0){\circle*{2}}
\put(60,0){\circle*{2}}
\put(70,0){\circle*{2}}
\put(80,0){\circle*{2}}
\put(85,-3){\line(0,1){2}}
\put(90,0){\circle*{2}}
\put(100,0){\circle*{2}}
\put(110,0){\circle*{2}}
\put(120,0){\circle*{2}}
\put(125,-3){\line(0,1){2}}
\put(130,0){\circle*{2}}
\put(140,0){\circle*{2}}
\put(150,0){\circle*{2}}
\put(160,0){\circle*{2}}
\qbezier(10,0)(40,27)(70,0)
\qbezier(20,0)(40,20)(60,0)
\qbezier(30,0)(40,12)(50,0)
\qbezier(80,0)(85,13)(90,0)
\qbezier(100,0)(130,27)(160,0)
\qbezier(110,0)(130,20)(150,0)
\qbezier(120,0)(130,12)(140,0)
\end{picture}
\end{equation}
for the $\varphi^4$ case. Observe that the corresponding graphs on the Euclidean setting would produce the same expressions as one finds in from the {\em ordinary}  of 1-particle-reducible graphs of ordinary field theory (all twisting factors are equal to 1 in this case),
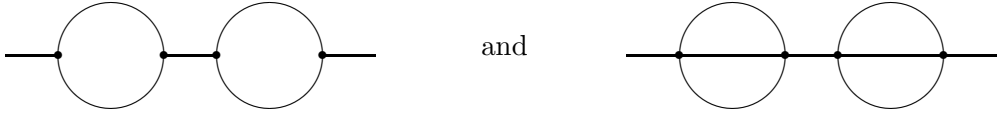
\[
\begin{picture}(200,50)(-30,0)
\put(0,20){\line(1,0){20}}
\put(20,20){\circle*{3}}
\put(40,20){\circle{140}}
\put(60,20){\circle*{3}}
\put(60,20){\line(1,0){20}}
\put(80,20){\circle*{3}}
\put(100,20){\circle{140}}
\put(120,20){\circle*{3}}
\put(120,20){\line(1,0){20}}
\end{picture}
\begin{picture}(30,50)
\put(10,20){$\mbox{and}$} 
\end{picture}
\begin{picture}(200,50)(-35,0)
\put(0,20){\line(1,0){140}}
\put(20,20){\circle*{3}}
\put(40,20){\circle{140}}
\put(60,20){\circle*{3}}
\put(80,20){\circle*{3}}
\put(100,20){\circle{140}}
\put(120,20){\circle*{3}}
\end{picture}
\]
We will first consider the graph (\ref{gr:ex_phi3}) from $\varphi^3$-theory in the Minkowskian setting. Its full twisting is 
\[
e^{-\frac \iu 2 (\tilde p_1 + \tilde p_2)\theta (k-\tilde p) } \ 
e^{-\frac \iu 2 (\tilde p_3 + \tilde p_4)\theta (k^\prime + \tilde p) }
\]
where $k$ and $k^\prime$ denote the external momenta, $\tilde p$ denotes the momentum  corresponding to the edge connecting the second  and the third vertex,  and $\tilde p_1, \tilde p_2$ and $\tilde p_3, \tilde p_4$ denote the pairs of inner momenta corresponding to the edges connecting the first and second, and the third and fourth vertex, respectively. 

By virtue of the fact that $\ra p_1+ \ra p_2=- \ra k= \ra p$, $\ra p_3 + \ra p_4 = \ra k^\prime$ and $k=-k^\prime$ in the adiabatic limit, we simplify the twisting as follows,
\[
e^{-\iu (\omega_{\ra p_1} + \omega_{\ra p_2}) (\theta k)_0  -\iu (k_0-\omega_{\ra k})(\theta k)_0 +\iu (\omega_{\ra p_3} + \omega_{\ra p_4}) (\theta k)_0 +\iu (k_0-\omega_{\ra k})(\theta k)_0 } 
 = 
e^{-\iu (\omega_{\ra p_1} + \omega_{\ra p_2}) (\theta k)_0   +\iu (\omega_{\ra p_3} + \omega_{\ra p_4}) (\theta k)_0  } 
\]
Choosing a time-ordering, we thus find the formal expressions 
\beqa
&&\int  g(x_1)\cdots g(x_4) \ e^{-\iu k x_1} \, e^{-\iu k^\prime x_4}\; 
  \tau(x_{1,0}-x_{2,0}) \, \tau(x_{2,0}-x_{3,0}) \, \tau(x_{3,0}-x_{4,0}) 
\ \Delta_+(x_2-x_3)
\\&& \qquad 
 \Delta^s_+(x_{1,0}-x_{2,0} + (\theta k)_0, \ra x_1 - \ra x_2)
\ 
 \Delta^s_+(x_{3,0}-x_{4,0} - (\theta k)_0, \ra x_3 - \ra x_4)
 \   dx_1 \cdots dx_4
\eeqa
with $s=2$ for the graph  (\ref{gr:ex_phi3}) from $\varphi^3$-theory,
 and $s=3$ for the graph (\ref{gr:ex_phi4}) from $\varphi^4$-theory.

Now,  even before the adiabatic limit is performed\footnote{The explicit form of the analytic expression corresponding to the graph (\ref{gr:ex_phi3}) in momentum space in the adiabatic limit can be found in appendix~\ref{app:calc_ex_phi3}.},  these two expressions in general no longer make sense as distributions on $\R^{2d+1}$, but instead require renormalization:

{\prop\label{prop:ill} \rm{ The formal integral kernel
\[
\tau(x_0) \tau(y_0) \ (t_{-a}\Delta_+^s(x)) \ ( t_{a}\Delta_+^s(y)) \ , \qquad k=2,3
\]
defines a distribution on $\dot \R^{2d+1}$. Its singular order in 0 is  $2d-9$ for $s=2$ and $4d -11$ for $s=3$, so the extension to $\R^{2d +1}$ is not unique in $d\geq 6$ dimensions, for both $s=2$ and $3$, and in $d= 4$ dimensions, it is not unique for $s=3$   ($d$ assumed even). 
}}

\proof{We collect the wavefront sets of the respective distributions on $\R^{2d +1}$
in a table, including  the information on the distributions' `arguments'; the last entry is the parameter with respect to which we take the translates $S$ and $T$:
\[
\begin{array}{r|ccccc|ccccc}
& \R & \R^{d-1} &  \R & \R^{d-1} &\R 
& \R & \R^{d-1} &  \R & \R^{d-1} &\R 
\\\hline 
\tau (x_0) & 0& \ra x & y_0& \ra y & a & \lambda & 0 & 0 &0&0
\\\hline 
\tau (y_0) &  x_0 & \ra x & 0 & \ra y& a& 0&0& \rho  & 0 & 0
\\
\hline
S^*\Delta_+^s (x) & x_0&0&y_0&\ra y&-x_0&p_0 & \ra p &0&0& p_0
\\
&x_0&\ra x \neq 0 &y_0&\ra y&\pm |\ra x| -x_0 & \mu |\ra x|& \mp \mu \ra x& 0&0&\mu |\ra x|
\\
\hline T^*\Delta_+^s (y) &x_0&\ra x& y_0&0&y_0&0&0 & k_0 & \ra k & -k_0
\\ 
&x_0&\ra x & y_0 & \ra y \neq 0 &y_0 \mp |\ra y| & 0&0&\nu |\ra y|& \mp \nu \ra y& - \nu |\ra y|
\end{array}
\]
Unless otherwise indicated, the entries on the left hand side (position space) are elements from all of $\R^{2d +1}$, while for the momenta we have the following constraints:
\[
\lambda \neq 0\ , \quad \rho \neq 0\ , \quad p_0 > |\ra p| \geq 0\ , \quad \mu>0 \ , 
\quad k_0 > |\ra k| \geq 0 \ , \quad \nu > 0
\]
From this table, we conclude that H\"ormander's criterion is satisfied for any two- and three-fold products  of these distributions on all of $\R^{2d+1}$. The product of all four of them (which is what is called the superficial divergence in renormalization theory), on the other hand, satisfies H\"ormander's criterion only outside $0$, while indeed, for $(x_0,\ra x, y_0,\ra y,a)=0$, the respective momenta can add up to 0: choose $\lambda=-p_0$, $\ra p=0$, $\rho=-k_0$, $\ra k=0$, and $p_0=k_0$.

The scaling degree of this divergence is therefore twice the scaling degree of the product of $\tau$ and $\Delta_+^s$, $s=2,3$ in 0. Hence for $s=2$ it is $4(d-2)$ -- but now, for a distribution on $\dot \R^{2d+1}$, so the singular order of the distribution is $4(d-2)-(2d+1)= 2d-9$ which is $-1$ for $d=4$ and positive for dimension $d \geq 6$. Therefore, in 4 dimensions, the distribution still makes sense on all of $\R^{2d+1}$, but requires to be renormalized for $d\geq 6$. 

For $s=3$, the scaling degree is twice $3(d-2)$, so the singular order is $6(d-2)-(2d+1)=4d-13$ which is positive for $d\geq 4$, such that this graph always  requires renormalization. 
}

It follows that in dimensions $d \geq 6$, the graph (\ref{gr:ex_phi3}) is ill-defined due to the ultraviolet divergence of powers of $\tau \Delta_+^s$ in 0, and that the graph  (\ref{gr:ex_phi4}) is ill-defined already for dimension $d\geq 4$. Recalling that $a=\lambda^2 k_2$, where $k=(k_0,\ra k)\in \R^{d+1}$ is the external momentum, we see that these divergences are (partly) assumed in the infrared regime, at small momentum. This means that we have found a mixing of divergences. It does not seem to be possibe to renormalize this divergence locally. 

Observe that multiple insertions of the fish graph ($\varphi^3$-graph) in $d=4$ do not increase the scaling degree: a straightforward extension of the proposition's proof reveals that the distribution corresponding to $k$ fish graphs lined up as in (\ref{gr:ex_phi3}) is defined on $\dot \R^{kd+1}$ and in 0 has singular order $k\cdot 2(d-2) - (kd+1) 
=-1$ for $d=4$.  However, in higher dimensions, and for the setting sun graph already in $d=4$, arbitrarily high scaling degrees occur by multiple insertions.

Last not least, it should be noted that contrary to the mixing in the Euclidean setting, this problem occurs already before we perform the adiabatic limit (although, admittedly, we used the adiabatic limit to simplify the twisting).

\section{Conclusion}

A mixing of ultraviolet and infrared divergences has been found in the Hamiltonian setting on the  noncommutative Moyal space (hyperbolic framework), settling a long-standing question.

The mechanism is different from the mixing found in theories with Euclidean signature. In fact, the graphs which exibit the problem in the Minkowskian realm, correspond to ordinary quantum field theory graphs in the Euclidean setting (no twistings appear there). Moreover, contrary to the mixing which was found in the Euclidean situation, the problem occurs even before the adiabatic limit is taken. 
 
As a consequence, it is very difficult to imagine a term which would, like the Grosse-Wulkenhaar-term~\cite{GrW} or the dressed propagator from~\cite{R} in the Euclidean setting, take care of these divergences. However, there is some indication that the perturbative expansion of the Euclidean Grosse-Wulkenhaar model might be Borel-summable, so it could be important  to try and find such a model also in the Minkowskian setting.

On the other hand, it seems to me that properties like the mixing of ultraviolet and infrared divergences could be an artefact of the canonical commutation relations we impose in Moyal space. It is to be hoped that more sophisticated models of quantum space time might have a better regularizing effect in quantum field theory.


\bigskip

{\bf Acknowledgement}

It is a pleasure to thank Jochen Zahn for many very helpful discussions and his detailed comments on this manuscript.

\medskip

Supported by the German Research Foundation (Deutsche Forschungsgemeinschaft (DFG)) via 
the Institutional Strategy of the University of G\"ottingen.


\begin{appendix}

\section{Rules}\label{app:rules}

To construct all \label{graphs} graphs at $k$-th order perturbation theory (in $\varphi^n$-theory), one writes down $k$ vertices, each of which consists of a row of $n$ dots, and then writes down all possible  ways to connect the dots (where to each dot at most one edge can be attached). Note that in this graphical language, there are no open edges: instead, the role of open edges in ordinary Feynman graphs is taken here by dots to which no edge is attached.

The rules how to recover the analytic expressions of the perturbative expansion from these graphs were derived e.g. in~\cite{filk, bahns_diss, sibold}. Here, they are given in a form that is particularly suited for the purposes of this paper. 

\subsection*{Modified Feynman Rules}

\begin{enumerate}
\item Label the vertices $1,\dots , k$.
\item \label{modFeyn_Schwinger} For any pair of dots, say the $r^{th}$ at vertex $ i$ and the $s^{\rm th}$ at vertex $j$ with $i<j$, which are connected by an edge, write down 
\[
\frac 1 {
p_{i, (r)}^2 +m^2} e^{- \iu  p_{i,(r)}(x_i-x_j)}  \qquad  \mbox{ where } p_{i,(r)} \in \R^d
\]
and for any pair of dots,  say the $r^{th}$  and the $s^{\rm th}$ at the {\em same} vertex $i$, with $r <s$, which are connected by an edge, write down 
\[
\frac 1 {
p_{i,(r)}^2+m^2} 
\]
\item For any dot, say the $r^{\rm th}$ dot of the vertex $i$, to which no edge is attached, write down an exponential 
\[
e^{-\iu p_{i,(r)} x_i}
\] 
\item \label{modFeyn_replace} Write down the twisting factor for each of the vertices with momenta 
$p_{i,(1)},\dots, p_{i,(n)}$ for vertex $i$, but taking into account the edges:
\begin{enumerate}
\item If an edge connects the $r^{\rm th}$ dot of vertex $ i$ with the $s^{\rm th}$ dot of vertex $j$ where $i<j$, then use $-p_{i,(r)}$ instead of $p_{j,(s)}$ at vertex $j$.
\item If an edge connects the $r^{\rm th}$ dot with the $s^{\rm th}$ dot of the {\em same} vertex $i$ where $r<s$, then use $-p_{i,(r)}$ instead of $p_{i,(s)}$ at vertex $i$.
\end{enumerate}
\item For each vertex $i$ write down a testfunction $g(x_i)$. 
Integrate over the vertices and over all those momenta which belong to an edge.
\end{enumerate}

\subsection*{Hamiltonian Rules}

\begin{enumerate}
\item Label the vertices $1,\dots , k$.
\item Pick a time-ordering and write down the appropriate Heaviside functions $\tau$.
The ordering of the vertices below now refers to this time ordering: if $x_{i,0}$ is later than $x_{j,0}$, we have $i>j$.
\item For any pair of dots, say the $r^{th}$ at vertex $ i$ and the $s^{\rm th}$ at vertex $j$ with $i>j$, which are connected by an edge, write down 
\[
\frac 1 {2\omega_{\ra p_{i, (r)}}}\  e^{- \iu  \tilde p_{i,(r)}(x_i-x_j)}  \qquad  \mbox{ where } \tilde p_{i,(r)}= (\omega_{\ra p_{i, (r)}},\ra p_{i, (r)}) \in \R^d
\]
Multiply with the appropriate Heaviside function $\tau(x_{i,0}-x_{j,0})$, if it is not yet present in the expression -- in which case, of course, the relative time ordering between $i$ and $j$ was deduced from combining the restrictions given by other  Heaviside functions.

For any pair of dots,  say the $r^{th}$  and the $s^{\rm th}$ at the {\em same} vertex $i$, with $r <s$, which are connected by an edge, write down 
\[
\frac 1 {2\omega_{\ra p_{i, (r)}}} 
\]
Observe that this latter rule means that we do not employ the systematics of (quasiplanar or ordinary) Wick products. 
\item For any dot, say the $r^{\rm th}$ dot of the vertex $i$, to which no edge is attached, write down an exponential 
\[
e^{-\iu p_{i,(r)} x_i}
\] 
\item \label{Ham_replace} Write down the twisting factor for each of the vertices with momenta 
$p_{i,(1)},\dots, p_{i,(n)}$ for vertex $i$, but taking into account the edges:
\begin{enumerate}
\item If an edge connects the $r^{\rm th}$ dot of vertex $ i$ with the $s^{\rm th}$ dot of vertex $j$ where $i<j$, then use $-\tilde p_{i,(r)}$ instead of $p_{j,(s)}$ at vertex $j$ and $\tilde p_{i,(r)}$ instead of $p_{i,(r)}$ at vertex $i$.
\item If an edge connects the $r^{\rm th}$ dot with the $s^{\rm th}$ dot of the {\em same} vertex $i$ where $r<s$, then use $-\tilde p_{i,(r)}$ instead of $p_{i,(s)}$ and $\tilde p_{i,(r)}$ instead of $p_{i,(r)}$ at vertex $i$.
\end{enumerate}
\item For each vertex $i$ write down a testfunction $g(x_i)$. 
Integrate over the vertices and over the spatial part of  all those momenta which belong to an edge.
\end{enumerate}

Observe: For $\theta =0$, we recover the ordinary hyperbolic Feynman rules from the ones above by making use of the equality
\[
G_F(x)= \tau(x_0) \Delta_+(x) +  \tau(-x_0) \Delta_+(-x)
\]
in the sense of distributions. The (at first sight possibly strange) third rule above guarantees that all necessary Heaviside functions  appear.


\section{Renormalization/Extension of distributions}~\label{app:renorm} 

Let $u \in \mc D^\prime(\dot R^n)$ with scaling degree $m$ such that its singular order $\sigma=m-n$ is non-negative. Choose a projection from $\mc D(\R^n)$ to the space of testfunctions which vanish in 0 with order $\sigma$,
\[
P_{\sigma,w}(g)(x)=g(x)-w(x)\sum_{|\alpha|\leq \sigma} \frac{x^{\alpha}
(-1)^{|\alpha|}
}{\alpha!} 
\ \delta^{(\alpha)}(g)
\]
where the testfunction $w$ must be equal to $1$ on a neighbourhood of $0$, but otherwise can be chosen at will. Given such a projection, the corresponding extensions $u_R \in \mc D^\prime(\R^n)$ of $u$  are given by
\[
u_R(g)= u(P_{\sigma,w}g) + \sum_{|\alpha|\leq \sigma}\frac{(-1)^{|\alpha|}\,c_\alpha}{\alpha!} \ \delta^{(\alpha)} (g)
\]
with constants $c_\alpha$, which parametrize the freedom we have to fix extension's value in the testfunctions $g(x):=w(x)x^{\alpha}$, $|\alpha|\leq \sigma$ (observe that $P_{\sigma,w}(x^\alpha w) = 0$ for $|\alpha|\leq \sigma$, so $u_R(x^\alpha w)= (-1)^{|\alpha|}\,c_\alpha$). These terms are what is called counterterms in the physics literature.

Example: On $\dot \R^4$, 
the singular order of $G_F^2$ is $0$, so we have
\[
\big(G_F^2\big)_R (g) = G_F^2 \big( g - g(0)\,w \big) 
\]
where we have set the constant $c_0=0$. In the language of physics, a choice different from 0 corresponds  to a finite mass renormalization.

The scaling degrees and counterterms for $\varphi^4$-theory (in the hyperbolic setting) have been calculated explicitly up to third order perturbation theory in~\cite{pinter}.


\section{Momentum space}\label{app:calc_ex_phi3}

In momentum space, in the adiabatic limit, the expression corresponding to the graph (\ref{gr:ex_phi3}) is (up to numerical factors),
\beqa
&& \delta^{(4)}(q_1+q_2) \ \frac 1 {q_{1,0}+ \omshi q 1 {}  - i \epsilon} \ \frac 1 {\omshi q 1 {}}
\int 
\frac 1 {\omsh k + \omega_{{\ra k} + {\ra q}_1} + q_{1,0} - i \epsilon} 
\ \frac 1 {\omsh p + \omega_{{\ra p} + {\ra q}_1} + q_{1,0} - i \epsilon} 
\ \times 
\\
&& \qquad \qquad \times \  e^{-i\big(\omsh k + \omega_{{\ra k} + {\ra q}_1} - \omsh p - \omega_{{\ra p} + {\ra q}_1} ) (\sigma q_1)_0 }  \; 
\frac 1 {\omsh k 
\; \omega_{{\ra k} + {\ra q}_1} 
 \; \omsh p
\; \omega_{{\ra p} + {\ra q}_1} }  \ d {\ra k} d {\ra p} 
\eeqa
Without the twisting, the integral clearly diverges for $d\geq 4$. In $d=4$,  this is the usual commutative graph with its two {\em ultraviolet} logarithmic divergences. So, the naive power counting gives a superficial degree of divergence 0 (while treating the above also as a distribution in the component $q_2 \in \R$, one finds $-1$ for the power counting degree of divergence in $d=4$, in accordance with Proposition~\ref{prop:ill}).
Observe that due to the relative sign, also when the twisting is present, it has no regularizing effect for $\{ {\ra k} = {\ra p} \} \subset \R^6$.  It is difficult to understand these divergences in terms of such a formal integral -- the language  of wavefront sets and distributions is much more appropriate.

\end{appendix}


\end{document}